\documentclass[11pt,oneside]{article}

\usepackage{algorithm2e, amsfonts, amsmath, amssymb, amsthm, bbm, booktabs, enumitem, graphicx, natbib, rotating, url, xcolor} 
\usepackage[noend]{algpseudocode}
\usepackage[english]{babel}
\usepackage[a4paper]{geometry}
\usepackage[colorlinks=true, citecolor=blue, urlcolor=magenta, linkcolor=blue]{hyperref}
\usepackage{tcolorbox}

\newcommand{\green}{\color{green}}

\newcommand{\cP}{\mathcal{P}}
\newcommand{\dd}{\, \mathrm{d}}
\newcommand{\E}{\mathbb{E}}
\newcommand{\hsp}{\hspace{0.2mm}}
\newcommand{\myR}{\mathsf{R}}
\newcommand{\myS}{\mathsf{S}}
\newcommand{\one}[1]{\ensuremath{\mathbbm{1}}(#1)}

\begin{document}

\strut\vspace{-3.6cm}
\begin{center}\begin{tcolorbox}[width=0.82\textwidth, colframe=teal, colback=cyan, coltitle=white, title=\textsf{\textbf{Note}}]
    \sffamily
    This post-print corrects a minor typographic error in \textbf{Algorithm \ref{alg:PAV}} of the \href{https://doi.org/10.1785/0220240209}{published version}: three instances of $x$ were intended to be $y$.
\end{tcolorbox}\end{center}

\bigskip

\begin{center}
\bf \Large 
Enhancing the statistical evaluation of earthquake forecasts -- An application to Italy
\end{center}

\centerline{\bf Jonas R. Brehmer}
\centerline{Heidelberg Institute for Theoretical Studies}
\medskip

\centerline{\textbf{Kristof Kraus} and \textbf{Tilmann Gneiting}}
\centerline{Heidelberg Institute for Theoretical Studies}
\centerline{Karlsruhe Institute of Technology (KIT)}
\medskip

\centerline{\textbf{Marcus Herrmann} and \textbf{Warner Marzocchi}}
\centerline{University of Naples, Federico II}

\bigskip

\centerline{\today}

\medskip

\begin{abstract} 

Testing earthquake forecasts is essential to obtain scientific information on forecasting models and sufficient credibility for societal usage.  We aim at enhancing the testing phase proposed by the Collaboratory for the Study of Earthquake Predictability \citep[CSEP,][]{Schoretal2018} with new statistical methods supported by mathematical theory.  To demonstrate their applicability, we evaluate three short-term forecasting models that were submitted to the CSEP-Italy experiment, and two ensemble models thereof.  The models produce weekly overlapping forecasts for the expected number of M4+ earthquakes in a collection of grid cells.  We compare the models' forecasts using consistent scoring functions for means or expectations, which are widely used and theoretically principled tools for forecast evaluation.  We further discuss and demonstrate their connection to CSEP-style earthquake likelihood model testing, and specifically suggest an improvement of the T-test.  Then, using tools from isotonic regression, we investigate forecast reliability and apply score decompositions in terms of calibration and discrimination.  Our results show where and how models outperform their competitors and reveal a substantial lack of calibration for various models.  The proposed methods also apply to full-distribution (e.\,g., catalog-based) forecasts, without requiring Poisson distributions or making any other type of parametric assumption. 

\end{abstract}

\section{Introduction}

Earthquake forecasting has broad, manifold impact: It demarcates the interface between seismology and society, being the basic scientific component of any sound seismic risk reduction strategy; simultaneously, it is the ultimate scientific challenge for seismologists, because it allows testing different hypotheses about earthquake occurrence processes. 

Owing to the physical complexity of earthquake occurrence and our lack of knowledge, earthquake forecasts are unavoidably probabilistic, meaning they provide the full or parts of the predictive distribution of future seismicity.  To be of scientific and practical use, the earthquake forecasting enterprise must be intimately linked to a solid testing phase, which allows seismologists to evaluate the reliability and predictive performance of the forecasts and give them the necessary credibility they need for societal purposes.  Enhancing the testing phase of earthquake forecasts is the main motivation of this paper. 

Presently, several flavors of physics-based \citep[e.g.,][]{cattania_forecasting_2018, mancini_improving_2019, sharma_is_2020, dahm_coulomb_2022} or statistical \citep[e.g.,][]{Gerstetal2005, Falconeetal2010, LombMarz2010, Woessneretal2010, bayliss_datadriven_2020} models exist that issue such forecasts.  To date, the most common methods to evaluate the resulting model output are set by the \textit{Collaboratory for the Study of Earthquake Predictability}~(CSEP) and rely on standardized prospective tests.  Prospective tests only use data gathered after the model was submitted to an experiment.  CSEP is a continuously evolving enterprise, both in the format of the forecasts and in the set of tests.  To preserve transparency and reproducibility, CSEP maintains the original models' forecasts and related data to allow any scientist to re-run the experiment with, e.g., an updated set of tests.  A key part of the CSEP testing strategy is an approach that can be broadly termed \textit{earthquake likelihood model testing}\/ \citep{KaganJack1995, Schoretal2007}.  In its original format, it partitions the testing region into grid cells and assigns a \textit{log likelihood}\/ to each forecast, which comprises Poisson log likelihoods for observed counts of earthquakes.  Although similar in form, the approach differs in crucial aspects from widely used statistical out-of-sample forecast evaluation methods developed for count data in other scientific fields \citep{czado2009, kolassa2016}.

In this paper we aim at providing additional statistical procedures that are deeply rooted in probability theory to evaluate and compare earthquake forecasts.  Specifically, we build on recent advances in \citet{Brehmetal2021} and illustrate how CSEP's log likelihood-based testing method is related to consistent scoring (or loss) functions, which are well-established tools from decision theory to compare forecast performance \citep{GneitRaft2007, Gneit2011}.  Furthermore, we introduce methods proposed by \citet{GneitResin2022} to investigate a model's reliability, and we apply score decompositions in terms of calibration and discrimination capabilities.  We use the newly proposed tools to compare and analyze five forecasting models regarding the operational earthquake forecasting (OEF) system in Italy \citep{Marzocchietal2014}.  The final goal is to propose additional mathematical resources that may make the CSEP testing phase stronger.

Terminology and technical settings matter, and to create an effective bridge between different fields, here seismology and mathematics, we need to pay attention to fundamental concepts and terminology of the mathematical theory used, starting with the details of the forecast format.  Specifically, we consider forecasts of the number of earthquakes in a given space--time--magnitude bin.  As argued by \citet{Nandanetal2019}, forecasting models ought to embrace the full distribution of the number of earthquakes.  However, a full-distribution forecast can always be reduced to its mean or expected value, and it is this simplified format on which we focus here.  Thus, the forecast is a nonnegative real number, which represents a mean or expected count, and the respective outcome is the observed count, which is a nonnegative integer.  Evidently, short-term earthquake forecasts operate within a \textit{low probability}\/ \citep{Jordanetal2011} or \textit{low count}\/ environment, where an overwhelming majority of the observed counts is zero.  \citet{Serafini2022} and \citet{Spassiani2023} consider an even simpler format, namely that of probability forecasts for the binary event of at least one target earthquake occurring in any given space--time--magnitude bin, to which we also relate.  The setting of single-valued forecasts in the form of an expected count or number, on which we focus in this paper, fits current practice at CSEP and allows for comprehensive comparisons.

\begin{figure}[tb]
\centering
\includegraphics[scale=1.0]{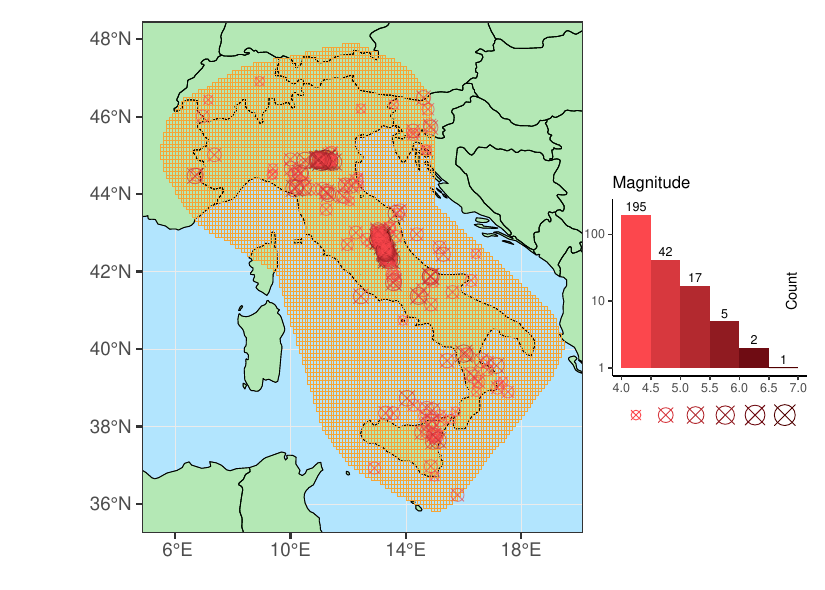}
 \caption{Left: Forecast region of OEF-Italy (orange grid, 8993 grid cells, which corresponds to the testing region of the Italian CSEP experiment) and locations of observed M4+ target earthquakes (crossed circles) between 2005 April 16 and 2020 May 20; Right: Logarithmic bar plot of earthquake magnitudes.  For similar displays, see Figure~1 of \citet{HM2023}, Figure~1 of \citet{Spassiani2023}, and Figure~2 of \citet{Brehmetal2021}.
\label{fig:testing_region}}
\end{figure}

Figure~\ref{fig:testing_region} depicts the Italian CSEP testing region in which the five models produce forecasts in the form of the expected number of earthquakes in space--time--magnitude bins.  We term the first two models as LM \citep{LombMarz2010} and FCM \citep{Falconeetal2010}.  Both are ETAS models, but with customized structure and calibration choices.  The third model, termed LG \citep{LG2003, Woessneretal2010}, is based on the short-term earthquake probability (STEP) model \citep{Gerstetal2005}.  These three models were evaluated in the Italian CSEP experiment \citep[][providing one-day ahead forecasts]{Taronietal2018}.  The fourth model, SMA, is a weighted-average ensemble of the previous three models and used in OEF-Italy \citep{Marzocchietal2014}; its weights are proportional to the inverse of the log likelihood of the observed data following the score model averaging (SMA) rule and are continuously updated with new observations.  We additionally include a further weighted-average ensemble of the three candidates, termed logistic regression-based weighted average (LRWA) \citep{HM2023}, where the weighting is based on logistic regression (specifically, the variant fit to M3+ earthquakes with fitting scheme \#2, see their Appendix~C). 

The models issue forecasts in the form of the expected number of earthquakes with magnitudes equal to or larger than four (M4+, the target threshold) in the next seven days for each of the 8993 spatial grid cells of the testing region.  On any given day, the forecast is a collection of 8993 positive values which we denote via $x_{c,t}^{(\hsp j)}$, where $j$ represents the model, $c$ the grid cell, and $t$ the day.  To evaluate these forecasts, we consider M4+ target earthquakes that occurred between 2005 April 16 and 2020 May 26 in the testing region.  This yields 5520 days in total and results in 5514 days of forecast--observation pairs.  The observations $y_{c,t}$ represent the number of M4+ target earthquakes in grid cell $c$ during the seven-day period starting at $t$, and become available after day $t + 6$.

\begin{figure}[p]
\centering
\includegraphics[scale=0.97]{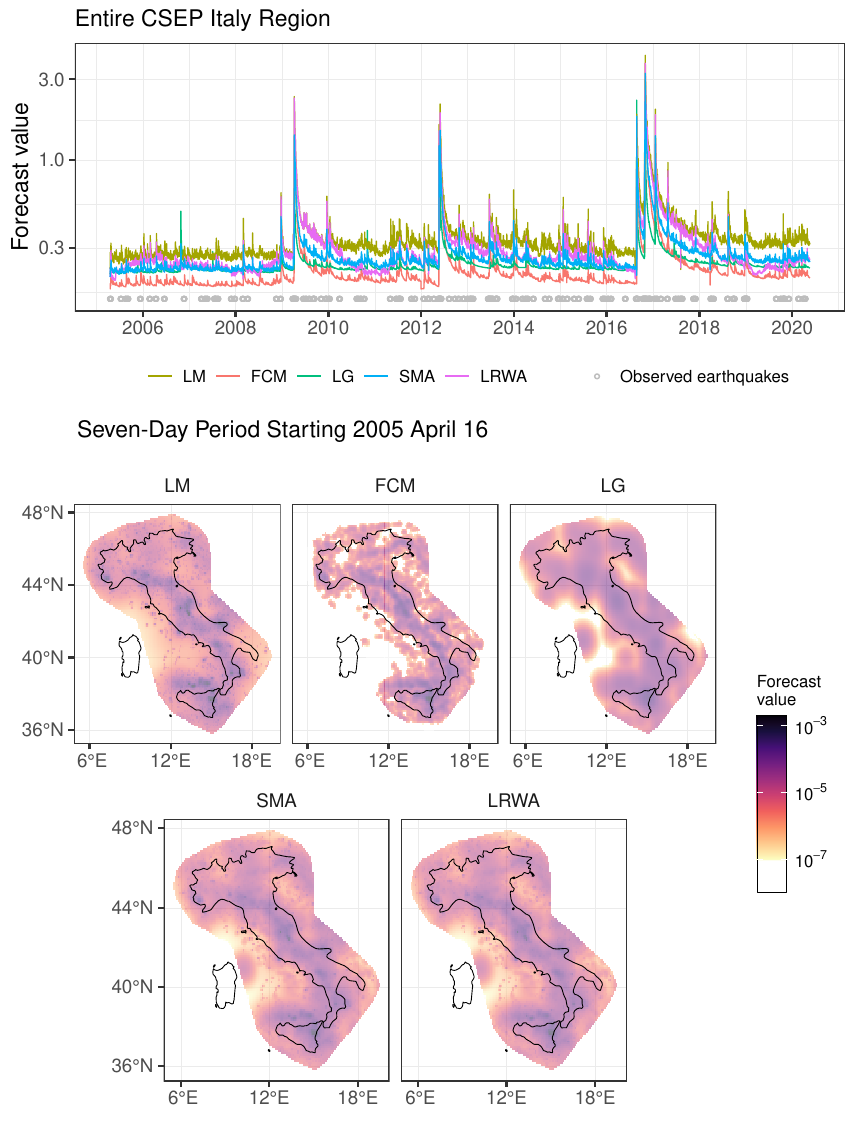}
\caption{Expected number forecasts of the five models.  Top: temporal evolution, aggregated over the testing region for each day; Bottom:  spatial distribution for the initial seven-day period in OEF-Italy, with forecast values below $10^{-7}$ represented in white.  
\label{fig:forecasts}}
\end{figure}

The models issue forecasts at 00:00 on each day and immediately after M3.5+ earthquakes.  We do not consider the latter (intra-day) runs and thus obtain an equally spaced sequence of forecasts in time.  A crucial benefit of the expected counts format is that forecasts can be aggregated spatially,\footnote{The aggregation is simply by summation.  For example, the expected count for the entire CSEP testing region equals the sum of the expected counts for the 8993 grid cells.  In contrast, full-distribution forecasts do not allow for linear nor other straightforward types of aggregation from bins.} as visualized in the top panel of Figure~\ref{fig:forecasts}.  We see that the models vary in several dimensions.  The respective forecasts have different background levels (top, bottom) and feature distinct spatial smoothness (bottom), though they are strongly correlated with each other.  In the next two sections we focus on terminology and theoretical concepts, but also provide extensive analyses and demonstrations based on these forecasts.

\section{Comparative evaluation via consistent scoring functions}  \label{sec:scoring}

Consistent scoring functions are mathematically principled tools for the overall evaluation, comparison, and ranking of competing forecast models.  We take scoring functions to be negatively oriented penalties that a forecaster wishes to minimize.  In a nutshell, if the forecast posits an expected count $x$ and the count $y$ realizes, the penalty is $\myS(x,y)$.  In practice, scores are suitably averaged.  

For instance, in the setting of OEF-Italy, which spatially coincides with the testing region of the Italian CSEP experiment, there are $C = 8993$ grid cells and $T = 5514$ test days.  The \textit{total score}\/ of model $j$ over the testing region and testing period from Figure~\ref{fig:testing_region} is 
\begin{equation}  \label{eq:Sbar} 
\bar\myS^{(\hsp j)} = \frac{1}{T} \sum_{t = 1}^T \bar\myS_t^{(\hsp j)}, 
\end{equation}
where 
\begin{equation}   \label{eq:Sbar_t}
\bar\myS_t^{(\hsp j)} = \sum_{c = 1}^C \myS \! \left( x_{c,t}^{(\hsp j)} , y_{c,t} \right)
\end{equation}
is the spatially aggregated score for model $j$ on day $t$.  Considering all $C$ grid cells and $T$ days, the total score considers $n = C \times T = 8993 \times 5514$ individual forecast cases in OEF-Italy.  Note that we use summation rather than averaging to define the spatially aggregated score at \eqref{eq:Sbar_t}, as the summation yields numerical values that are convenient to report, compare, and interpret. 

A key question then is which scoring function ought to be used, and we emphasize that the choice needs to be mathematically consistent with the scope and format of the forecast, which represents the \textit{a priori}\/ mean or expected number of earthquakes.  Despite our focus on forecasts in the form of expected counts, it is instructive to begin the discussion in the setting of full-distribution forecasts. 

\subsection{Proper scoring rules for full-distribution forecasts}  \label{sec:proper}
 
To fix the idea, let $P$ be a full-distribution forecast for count data, such as the number of earthquakes in a space--time--magnitude bin.  Evidently, $P$ is a probability distribution on the set of the nonnegative integers.  We adopt notation from \citet{czado2009} and denote its probability mass function by $(p_k)_{k = 0}^\infty$ and the respective cumulative distribution function (CDF) by $(P_k)_{k = 0}^\infty$, where $P_k = \sum_{j = 0}^k p_j$ for any integer $k \geq 0$.  

In this setting, a scoring rule $\myR(P,y)$ assigns a penalty based on the full-distribution forecast $P$ and the corresponding observed count $y$.  The scoring rule $\myR$ ought to be such that 
\begin{equation} \label{eq:proper}
\E_{Y \sim \hsp Q} \hsp \hsp \myR(Q,Y) \: \leq \: \E_{Y \sim \hsp Q} \hsp \hsp \myR(P,Y), 
\end{equation}
where $\E$ denotes the expectation operator, $P$ and $Q$ are full-distribution forecasts, and $Y \sim Q$ specifies that $Y$ is a random variable with distribution $Q$.  Indeed, if $Q$ is the true (though typically unknown) distribution of the count, then the full-distribution forecast $Q$ ought to outperform any other forecast $P$ in terms of the expected value of the score.  A scoring rule for which the expectation inequality in \eqref{eq:proper} is true for all $P, Q$ in a class $\cP$ of probability distributions is said to be \textit{proper}\/ relative to $\cP$.  If equality in \eqref{eq:proper} occurs only when $P = Q$, the scoring rule is \textit{strictly proper}.  Propriety is an essential characteristic of a scoring rule which encourages honest and coherent forecasts \citep{GneitRaft2007} and ensures that appropriate and complete usage of information is rewarded \citep{holzmann2014}.

Key examples of proper scoring rules for count data include the \textit{logarithmic}\/ score, which is defined as
\begin{equation}  \label{eq:logarithmic}
\myR(P,y) \: = \: - \log p_y
\end{equation}
and depends on the forecast $P$ only through the probability mass $\, p_y$ at the observed count, and  the \textit{ranked probability}\/ score, 
\begin{equation}  \label{eq:rps}
\myR(P,y) 
\: = \: \sum_{k = 0}^y P_k^2 \: + \sum_{k = y + 1}^\infty (1-P_k)^2
\: = \: \E_{Y \sim \hsp P} \hsp \hsp |Y-y| - \frac{1}{2} \, \E_{Y \sim \hsp P} \hsp \hsp |Y-Y'|,
\end{equation}
where $Y$ and $Y'$ are independent copies of a random variable with distribution $P$.  The representation at right in \eqref{eq:rps} holds whenever the distribution $P$ has a finite (as opposed to infinite) expected value, and it demonstrates that the ranked probability score can be interpreted in terms of (fractional) counts.  \citet{czado2009} and \citet{kolassa2016} provide detailed discussions of these and other proper scoring rules.

\subsection{Consistent scoring functions for expected counts}  \label{sec:consistent}

In the context of forecasts of a mean or expected count, consistency is the equivalent of propriety \citep{Savage1971, Gneit2011}.  Specifically, let $\cP$ be the class of probability distributions on the nonnegative integers with finite mean or expectation, and let $P \in \cP$ have expectation $\mu_P$.  The scoring function $\myS$ is \textit{consistent}\/ for forecasts of an expected count if
\begin{equation}   \label{eq:consistent}
\E_{Y \sim \hsp P} \hsp \hsp \myS(\mu_P, Y) \leq \E_{Y \sim \hsp P} \hsp\hsp \myS(x, Y)
\end{equation}
for all distributions $P \in \cP$ and nonnegative numbers $x$.  The scoring function is \textit{strictly consistent}\/ if equality in~\eqref{eq:consistent} implies $x = \mu_P$.  Consistency ensures that reporting the truth is an optimal strategy when forecasters are rewarded according to their realized scores, and strict consistency guarantees that the true expectation $\mu_P$ is the only minimizer of the expected score $\E_{Y \sim \hsp P} \hsp \hsp \myS(x,Y)$.  Hence, a forecast is preferred over its competitors if it achieves lower average scores.  To see the connection to strictly proper scoring rules, we note that if $\myS(x,y)$ is a consistent scoring function in the sense of Eq.~\eqref{eq:consistent} then $\myR(P,y) = \myS(\mu_P,y)$ is a proper scoring rule for full-distribution forecasts in the sense of Eq.~\eqref{eq:proper}.\footnote{Indeed, if $P, Q \in \cP$ then $\E_{Y \sim \hsp Q} \hsp \hsp \myR(Q,Y) = \E_{Y \sim \hsp Q} \hsp \hsp \myS(\mu_Q,Y) \leq \E_{Y \sim \hsp Q} \hsp \hsp \myS(\mu_P,Y) = \E_{Y \sim \hsp Q} \hsp \hsp \myR(P,Y)$, which demonstrates the claim.} 

Due to the condition~\eqref{eq:consistent} prescribed by consistency, it is an intricate question what consistent scoring functions are available for expected counts.  Summarizing results originally due to \citet{Savage1971} and reviewed by \citet{Gneit2011} and \citet{Brehmetal2021}, among others, the (strictly) consistent scoring functions for an expected count are of the form
\begin{equation}   \label{eq:Bregman}
\myS(x,y) = \phi(y) - \phi(x) - \phi'(x)(y - x) + h(y),
\end{equation}
where $x \geq 0$, $y \in \{ 0, 1, \ldots \}$, $\phi$ is a (strictly) convex function, $\phi'$ is a subgradient of $\phi$, and $h$ is an essentially arbitrary function.\footnote{A subgradient is a generalization of a derivative; whenever a derivative exists, the subgradient is unique and coincides with the derivative.  For further technical discussion we refer to Section~3.1 of \citet{Gneit2011}, Section~2 of \citet{Brehmetal2021}, and Appendix~\ref{app:Bregman} in this current paper.} 

The most prominent example of a strictly consistent scoring function for an expected value in the scientific literature is the \textit{quadratic}\/ scoring function 
\begin{equation}   \label{eq:quadratic}
\myS_\mathrm{quad}(x,y) = \left( x - y \right)^2,
\end{equation}
which arises from $\phi(x) = x^2$ and $h(y) = 0$ in \eqref{eq:Bregman}.  This function is symmetric in the sense that $\myS_\mathrm{quad}(x,y) = \myS_\mathrm{quad}(y,x)$.  Another valid choice of a convex function in~\eqref{eq:Bregman} is $\phi(x) = x \hsp (\log x - 1)$.  Taking $h(y) = - \phi(y)$, this yields the \textit{Poisson}\/ scoring function $\myS_\mathrm{pois}$ defined by
\begin{equation}   \label{eq:Poisson}
\myS_\mathrm{pois}(x,y) = x - y \log x
\end{equation}
for expected counts $x > 0$ and observed numbers $y \in \{ 0, 1, \ldots \}$ of target earthquakes, which also is strictly consistent.  The case $x = 0$ does not occur in typical practice, but can be handled by assigning $\myS_\mathrm{pois}(0,0) = 0$ and $\myS_\mathrm{pois}(0,y) = + \infty$ for $y \in \{ 1, 2, \ldots \}$.

For a connection to proper scoring rules, note that the logarithmic score at \eqref{eq:logarithmic} recovers the Poisson score at \eqref{eq:Poisson} when $p_y$ is the probability mass of a Poisson distribution with rate $x > 0$, save for terms that depend on the outcome $y$ only.  In other words, $\myS_\mathrm{pois}(x,y)$ corresponds to the log likelihood of a Poisson distributed random variable and therefore is equivalent to the main evaluation function used in the earthquake likelihood model testing approach \citep{Schoretal2007}.  However, the connection is purely formal, and the use of the Poisson scoring function neither explicitly nor implicitly requires Poisson distributions or any other type of parametric assumptions, as pointed out by \citet{Brehmetal2021}.  Both the quadratic and the Poisson scoring function are members of the \textit{extended Patton family}\/, which is an extension to count data of the scoring functions proposed by \cite{Patton2011}.  For details see Appendix~\ref{app:Patton}.

\begin{table}[ptb]
\centering
\caption{Predictive performance of forecast models for OEF-Italy in terms of the average score $\bar\myS$ from \eqref{eq:Sbar} under the Poisson scoring function at \eqref{eq:Poisson} and the quadratic scoring function at \eqref{eq:quadratic}.  The best (lowest) score in each column is highlighted in green.  \label{tab:scores}}
\medskip
\begin{tabular}{lcc}
\toprule
Score & Poisson & Quadratic \\
\midrule
LM  &   2.71   &  \green{ 0.8407 } \\
FCM  &   2.80   &   0.8459  \\
LG  &   3.02   &   0.8465  \\
\midrule
SMA  &   2.73   &   0.8437  \\
LRWA  &  \green{ 2.69 }  &   0.8421  \\
\bottomrule
\end{tabular}
\end{table}

As Table~\ref{tab:scores} shows, the quadratic score and the Poisson score yield distinct rankings of the five models.  Under the quadratic scoring function, the LM model has the best (lowest) score; under the Poisson scoring function the LRWA model shows the lowest score.\footnote{Differences between Tables~\ref{tab:scores}--\ref{tab:IG} in this current paper and Table~1 in \citet{Brehmetal2021}, as well as between Table~\ref{tab:IG} and Table~1 in \citet{HM2023}, stem from a corrected spatial binning of earthquakes that occurred exactly on grid cell boundaries.  Details are available at \url{https://github.com/jbrehmer42/Earthquakes_Italy}.}  As there are many strictly consistent scoring functions, it is unclear whether conclusions are reasonably stable with respect to the choice of scoring function.  In investigating this stability, we rely on an alternative characterization of scoring functions from \citet{Ehmetal2016}.  To this end, we define the \textit{elementary}\/ scoring function $\myS_\theta$ with threshold parameter $\theta > 0$, namely, 
\begin{equation}   \label{eq:elementary}
\myS_\theta (x,y)
= \left\{ 
\begin{array}{ll} 
0,            & x, y \leq \theta \; \textrm{ or } \; x, y \geq \theta, \\
|y - \theta|, & \textrm{otherwise},
\end{array} 
\right. 
\end{equation}
which arises from the classical representation \eqref{eq:Bregman} by choosing the convex function $\phi(x) = \max(x - \theta, 0)$ and $h(y) = 0$.  Thus, $\myS_\theta$ assigns a score of zero if the expected number $x$ and the observed count $y$ lie on the same side of $\theta$, and a score that equals the distance between $y$ and $\theta$ otherwise.  We note that $\myS_\theta$ is a consistent, but not a strictly consistent, scoring function since $x = y$ is not the only minimizer of the score.

Remarkably, $\myS_\theta$ represents a cost--benefit scenario in a stylized decision problem with a decision threshold at $\theta$, where $\theta$ can be interpreted as the ratio of the socio-economic cost of a preventive measure under consideration, such as an evacuation, over the (monetized) benefit (per earthquake) of the measure, e.\,g., the number of saved human lives (per earthquake).  If the expected number of earthquakes $x$ lies below the threshold $\theta$, the induced optimal decision is to take no action; if $x$ is above $\theta$, the induced optimal decision is to apply the preventive measure.\footnote{Let $c$ be the socio-economic cost of the preventive measure, and let $b$ denote its monetized benefit per earthquake.  Then the measure ought to be undertaken if $bx > c$ or $x > c/b = \theta$.}  In this scenario $\myS_\theta(x,y)$ at \eqref{eq:elementary} is proportional to the difference of the cost or loss under the optimal decision induced by the forecast $x$ and the (hypothetical) cost under the optimal decision in hindsight \citep[Section~3.1]{Ehmetal2016}.  Notwithstanding the crude simplifications in the stylized decision problem, decision thresholds (only) take (very) small positive values in practical problems of short-term forecasts of damaging earthquakes.

The results of \citet{Ehmetal2016} imply that subject to modest regularity conditions, a consistent scoring function $\myS$ for an expected count admits a representation of the form
\begin{equation}   \label{eq:mixture}
\myS(x,y) = \int_0^{\infty} \myS_\theta (x, y) \: w(\theta) \dd \theta,
\end{equation}
where $w(\theta)$ is a non-negative weight function that assigns relevance to the threshold parameter $\theta > 0$ for the elementary scoring function $\myS_\theta$ from \eqref{eq:elementary}.  As noted by \citet[][p.~306]{Taggart2022}, the score is ``the weighted average of economic regret over user decision thresholds, where the weight emphasises those decision thresholds in the corresponding region of interest.''  The scoring function $\myS$ in \eqref{eq:mixture} is strictly consistent if the weight function $w(\theta)$ is strictly positive.  If we put $h(x) = 0$, then the convex function $\phi$ in the classical representation \eqref{eq:Bregman} and the weight function $w$ in the nearly equivalent, but more interpretable, representation \eqref{eq:mixture} relate to each other in simple ways.  Specifically, if $\phi$ admits a second derivative $\phi''$, then $w(\theta) = \phi''(\theta)$ for $\theta > 0$.  For example, the quadratic scoring function at \eqref{eq:quadratic} arises from the choice $\phi(x) = x^2$, which yields the uniform weight function $w(\theta) = \phi''(\theta) = 2$ for $\theta > 0$, so all values of the decision threshold $\theta$ are weighted equally.  The Poisson scoring function at \eqref{eq:Poisson} arises from the convex function $\phi (x) = x \hsp (\log x - 1)$, which yields the corresponding weight function $w(\theta) = \phi''(\theta) = \theta^{-1}$ for $\theta > 0$, so small positive values of the decision threshold $\theta$ receive dominant weights.  Importantly, this analysis demonstrates that in the setting of earthquake forecasting, the Poisson scoring function with its emphasis on small positive decision thresholds is more relevant than the quadratic scoring function.

To investigate how forecast performance depends on the choice of the consistent scoring function, we can plot a Murphy diagram.  In the original form proposed by \citet{Ehmetal2016}, the Murphy diagram plots the average score \eqref{eq:Sbar} under the elementary scoring function $\myS_\theta$ from \eqref{eq:elementary} for the model(s) at hand as a function of the threshold parameter $\theta$.   Here we adapt the Murphy diagram to the low probability and low count setting of earthquake forecasts and introduce the \textit{logarithmic Murphy diagram}, where we plot against $\log(\theta)$.  Conveniently, the integral under a model's logarithmic Murphy curve equals its average Poisson score.\footnote{The claim follows immediately from our discussion of the representation \eqref{eq:mixture} for the Poisson scoring function and the fact that $\int \bar\myS_\theta \dd(\log \theta) = \int \bar\myS_\theta \hsp \theta^{-1} \dd \theta$.  The average score \eqref{eq:Sbar} under the elementary scoring function $\myS_\theta$ at~\eqref{eq:elementary} is identically zero for sufficiently small and sufficiently large values of $\theta$, so the integral is guaranteed to be finite.}  If a model has a lower average elementary score than its competitors for every choice of the threshold parameter $\theta$, then it has superior forecast performance with respect to all strictly consistent scoring functions for the mean.  

\begin{figure}[tb]
\centering
\includegraphics[scale=1.0]{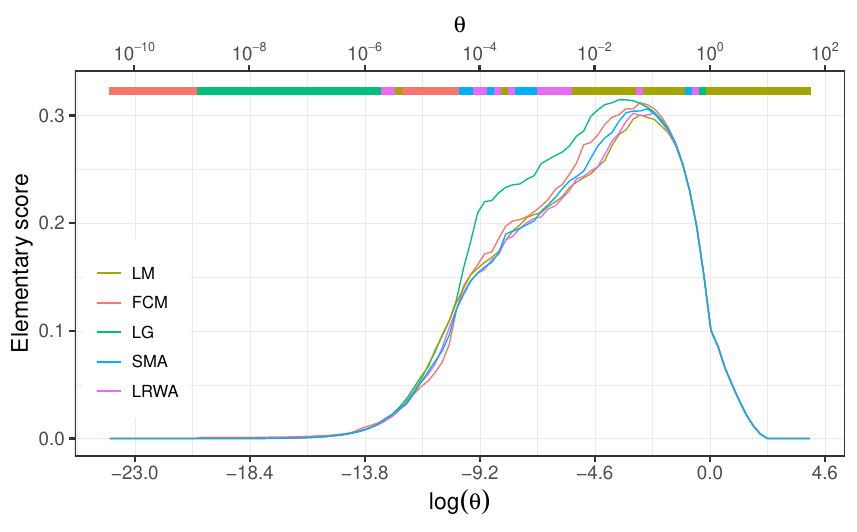}
\caption{Logarithmic Murphy diagram for the five forecast models.  Each curve plots a model's total elementary score $\bar\myS_\theta$ from \eqref{eq:Sbar} versus $\log \theta$.  Tickmarks at bottom indicate $\log \theta$; tickmarks at top show $\theta$.  The colored bar at top indicates the model with the lowest value of $\bar\myS_\theta$.  The integral under a model's curve equals the average Poisson score from Table~\ref{tab:scores}.  \label{fig:logMurphy}}
\end{figure}

Figure~\ref{fig:logMurphy} shows a logarithmic Murphy diagram for the five OEF-Italy forecast models.  It is interesting to observe that each of the five models has certain ranges of $\theta$ under which it performs best in terms of the average score under $\myS_\theta$.  The LM model shows outstanding performance at larger values of $\theta$ and thus has the smallest average score under the quadratic scoring function, which gives equal weight to all values of $\theta$.  The LRWA ensemble model shows excellent performance across all values of $\theta$ and has the smallest average score under the Poisson scoring function.  As we have argued, the Poisson scoring function emphasizes performance at small values of $\theta$ that societally are the most relevant, and a model's average Poisson score equals the area under the logarithmic Murphy curve.  

\subsection{The Diebold--Mariano test of equal predictive ability}  \label{sec:DM}

Worthy of note, scoring functions rank the forecast models' performance, but to make formal statistical inference, additional procedures and assumptions are required.  To check for statistically significant differences in forecast performance, we consider tests of the (null) hypothesis of equal predictive performance for two models in terms of a given scoring function $\myS$.  For this purpose, we make use of the Diebold--Mariano test, which is a carefully adapted Student's $t$-test \citep[][Section~3.1]{DiebMari1995, GK2014}.

\begin{figure}[p]
\centering
\includegraphics[scale=0.98]{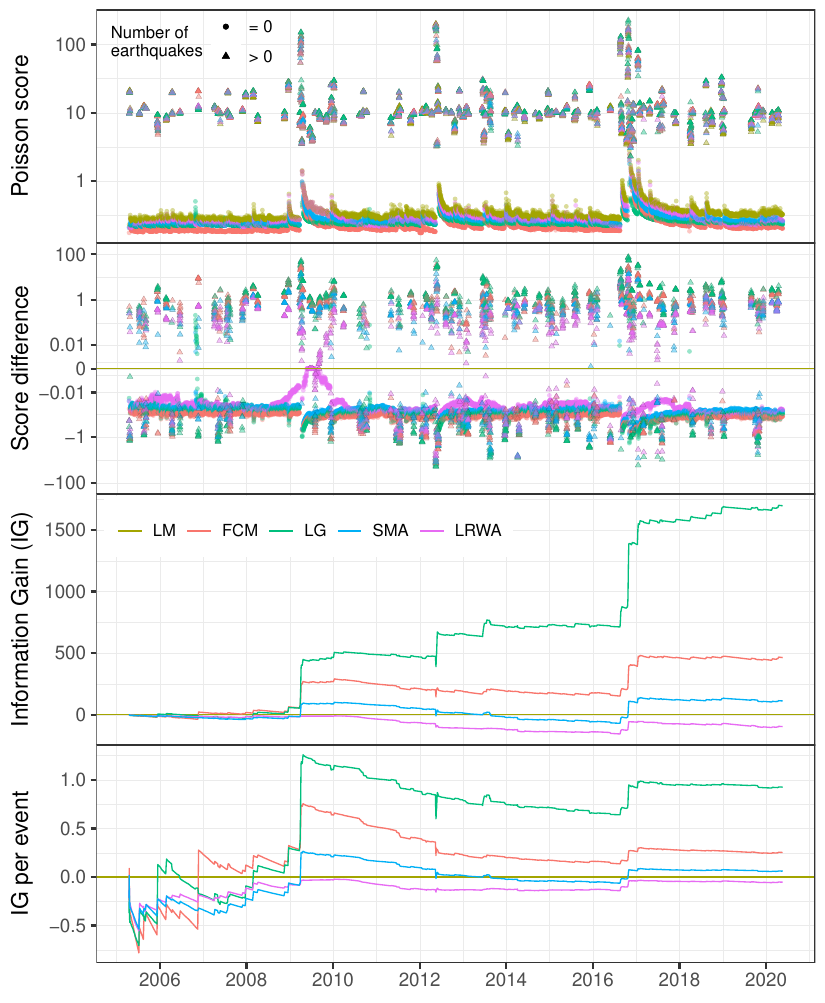}
\vspace{-2mm}
\caption{From top to bottom: Spatially aggregated daily Poisson score \eqref{eq:Sbar_t} for the five forecast models; Daily Poisson score difference relative to the LM model; Cumulative Poisson score difference or information gain (IG) of the LM model over the other models; Information gain per earthquake (IGPE) of the LM model over the other models.  In the first two panels, two different markers are used: triangles for periods with one or more M4+ target earthquakes, and circles otherwise.  Note the logarithmic scale in the upper two panels.  All quantities are negatively oriented, i.e., the smaller the better for the color-coded model.  For technical details see Appendix~\ref{app:figure}.  \label{fig:score.IG.IGPE}}
\end{figure}

Specifically, we consider the daily scores $\bar\myS_t^{(\hsp j)}$ and $\bar\myS_t^{(k)}$ of models $j$ and $k$ on day $t$ from \eqref{eq:Sbar_t}, which are obtained after spatial aggregation, with the individual scores for the grid cells summed up, and summarize forecast performance over the whole testing region for one seven-day forecast period.\footnote{Appendix~\ref{app:individual} concerns spatial displays of Poisson score differences, obtained after temporal aggregation.}   The top panel in Figure~\ref{fig:score.IG.IGPE} shows these daily scores based on the Poisson scoring function $\myS_\mathrm{pois}$ for the five forecast models.  It uses a logarithmic scale, because the daily scores are much larger for forecast periods $t$ that contain M4+ target earthquakes, corresponding to counts $y_{c,t} \geq 1$ for at least one grid cell $c$.  The LM model generally performs worst on forecast periods without M4+ earthquakes, as it possesses the highest daily scores for such $t$, and frequently performs best on forecast periods with M4+ earthquakes, when it possesses low scores.  

This is mirrored by the second panel in Figure~\ref{fig:score.IG.IGPE}, which shows the difference $\bar\myS_t^{(\hsp j)} - \bar\myS_t^{(k)}$ in the daily scores between the LM model ($k$), which we employ as reference throughout the paper, and its competitors ($\hsp j$), again using the Poisson scoring function.   In forecast periods with M4+ earthquakes, the majority of the score differences are positive, so the LM model attains the best scores and the FCM model the worst scores for these periods.  A look back at the top panel of Figure~\ref{fig:forecasts} provides a straightforward explanation: the LM model issues the highest expected count forecasts, whereas the FCM model issues the lowest forecasts.  The bottom two panels in Figure~\ref{fig:score.IG.IGPE} show the information gain (IG) and information gain per earthquake (IGPE) of the LM model over the other models.  As we prove analytically in Appendices~\ref{app:figure} and \ref{app:Poisson}, IG and IGPE are multiples of the cumulative, spatially aggregated Poisson score difference between the model at hand and the LM model.  However, while IG can grow indefinitely, IGPE is normalized by the cumulative number of observed target earthquakes and thus tends to stabilize over the course of the evaluation period.

To check for statistically significant differences in forecast performance, we apply the Diebold--Mariano test on the daily scores $\bar\myS_t^{(\hsp j)}$ from \eqref{eq:Sbar_t}, which operates under the null hypothesis that model $j$ and model $k$ perform equally well in terms of the scoring function $\myS$.\footnote{The Diebold--Mariano test applies not only under the Poisson score, but under any scoring rule or scoring function, such as those discussed by \citet{Lerch2017}, \citet{Serafini2022}, \citet{Brehmetal2021}, and \citet{Heinrich2024}, among others.}  If the null hypothesis is true, the test statistic
\begin{equation}  \label{eq:z}
z_{(\hsp j,k)} = \sqrt{T} \: \frac{\bar\myS^{(\hsp j)} - \bar\myS^{(k)}}{\hat{\sigma}_{(\hsp j,k)}}
\end{equation}
has a standard normal distribution in the limit as the number $T$ of testing times grows larger.\footnote{The standard normal distribution arises as the limit of the Student's $t_\nu$-distribution for an increasing number of degrees of freedom $\nu$.  The number $T$ of testing times needs to be sufficiently large for the limit to be applicable; typically, values of $T$ in the low thousands suffice.  Note the stark and pleasant contrast to the CSEP T-test: For the Diebold--Mariano test, we have theoretical guarantees of its validity (in the sense described in the next footnote) in sufficiently large samples.  For the CSEP T-test there are no theoretical guarantees available.  In fact, we are unaware of even a single setting in which the CSEP T-test yields the required uniformity of the $p$-value under the null hypothesis.}  In \eqref{eq:z}, $\hat{\sigma}_{(\hsp j,k)}^2$ is an estimate of the variance of the daily score differences that accounts for temporal dependencies and related effects \citep{DiebMari1995}, namely,  
\begin{equation}  \label{eq:sigma}
\hat\sigma_{(\hsp j,k)}^2 = \hat\gamma_{(\hsp j,k)}(0) + 2 \sum_{l = 1}^L \hat\gamma_{(\hsp j,k)}(l)
\end{equation}
where 
\[
\hat\gamma_{(\hsp j,k)}(l) = \frac{1}{T} \sum_{t = l + 1}^T (\bar\myS_t^{(\hsp j)} - \bar\myS_t^{(k)} - \bar{d}_{(\hsp j,k)}) (\bar\myS_{t-l}^{(\hsp j)} - \bar\myS_{t-l}^{(k)} - \bar{d}_{(\hsp j,k)}) \quad \text{and} \quad \bar{d}_{(\hsp j,k)} = \bar\myS^{(\hsp j)} - \bar\myS^{(k)}.
\]
The summation in the expression \eqref{eq:sigma} for $\hat\sigma_{(\hsp j,k)}^2$ is over the sample autocovariance $\hat\gamma_{(\hsp j,k)}(l)$ of the score differences at lags $l$ up to an integer value $L$ after which the score differentials can be considered uncorrelated.  If $L \geq 1$ then \eqref{eq:sigma} applies; if $L = 0$ then $\hat\sigma_{(\hsp j,k)}^2 = \hat\gamma_{(\hsp j,k)}(0)$.  For OEF-Italy, we set $L = 6$ due to the overlap of the seven-day test periods at lags up to six days.

A positive value of $z_{(\hsp j,k)}$ at \eqref{eq:z} arises from a smaller average score for model $k$; a negative value stems from a smaller average score for model $j$.  From $z_{(\hsp j,k)}$, a one-sided $p$ value can be derived, namely, 
\begin{equation}  \label{eq:p}
p = 1 - \Phi \! \left( z_{(\hsp j,k)} \right) \! ,  
\end{equation}
where $\Phi$ is the CDF of a standard normal random variable.  Under the null hypothesis of equal predictive ability in terms of the scoring function $\myS$ the $p$ value at \eqref{eq:p} is approximately uniformly distributed.  A small value of $p$ (typically, smaller than 0.01 or 0.05) allows for the rejection of the null hypothesis in favor of the alternative hypothesis of superior predictive ability of model $k$; a large value of $p$ (typically, larger than 0.95 or 0.99) suggests superior predictive ability of model $j$.

Table~\ref{tab:DM} shows the results for the five forecast models.  The Diebold--Mariano test rejects the null hypothesis of equal predictive ability in terms of the Poisson scoring functions for all pairs of models, except for the pairs LRWA and LM, LM and SMA, and LM and FCM.  The LRWA ensemble model shows the best performance in terms of the Poisson score, with score differences that are statistically significant, except for the comparison to the LM model. 

\begin{table}[ptb]
\centering
\caption{Diebold--Mariano test of the null hypothesis of equal predictive ability in terms of the Poisson scoring function for the five forecast models.  We show the respective total Poisson score (diagonal), the $z$ statistic (above diagonal), and the one-sided $p$ value (below diagonal).  Columns identify model $j$ and rows model $k$ in the $z$ statistic at \eqref{eq:z}.  \label{tab:DM}}
\medskip
\begin{tabular}{lccccc}
\toprule
& \begin{turn}{90} LRWA \hsp \end{turn}
& \begin{turn}{90} LM \hsp \end{turn}
& \begin{turn}{90} SMA \hsp \end{turn}
& \begin{turn}{90} FCM \hsp \end{turn}
& \begin{turn}{90} LG \hsp \end{turn} \\
\midrule
LRWA  &  \fbox{ 2.69 }  &   1.17   &   1.64   &   2.27   &   2.84  \\
LM  &   0.12   &  \fbox{ 2.71 }  &   0.60   &   1.56   &   2.45  \\
SMA  &   0.05   &   0.27   &  \fbox{ 2.73 }  &   2.46   &   3.05  \\
FCM  &   0.01   &   0.06   &   0.01   &  \fbox{ 2.80 }  &   2.83  \\
LG  &   0.00   &   0.01   &   0.00   &   0.00   &  \fbox{ 3.02 } \\
\bottomrule
\end{tabular}
\end{table}

\begin{figure}[ptb]
\centering
\includegraphics[scale=1.0]{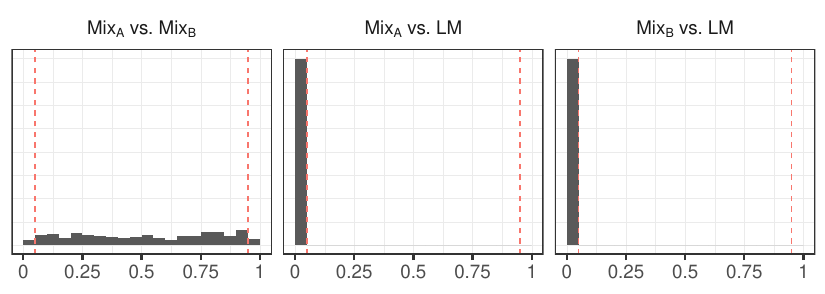}
\vspace{-2mm}
\caption{Histograms of $p$ values for Diebold--Mariano tests of equal predictive ability in terms of the Poisson scoring function for (Left) Mix$_A$ versus Mix$_B$; (Middle) Mix$_A$ versus LM; and (Right) Mix$_B$ versus LM, based on 400 replicates.  \label{fig:DM}}
\end{figure}

A valid test of significance\footnote{We use the term \textit{valid test}\/ in the sense of satisfying conditions (3.10) and (3.14) in \citet{Lehmann2022}, which conforms with typical practice throughout the scientific literature.  By Lemma 3.3.1 in \citet{Lehmann2022}, these conditions guarantee the uniformity of the $p$ value under the null hypothesis.} generates $p$ values that are uniform between 0 and 1 when the null hypothesis is true, and shows power by rejecting the null hypothesis with high probability when it is false \citep{Lehmann2022}.  We now report on a simulation experiment that demonstrates these desirable properties for the Diebold--Mariano test.  First, we consider two models for which the null hypothesis of equal predictive ability is true.  To achieve this, we define models Mix$_A$ and Mix$_B$ in the setting of OEF-Italy.  At any given time $t$, Mix$_A$ chooses randomly between the FCM and the LG forecast, with equal probability, whereas Mix$_B$ uses the other forecast.  Viewed as stochastic processes, Mix$_A$ and Mix$_B$ have equal predictive ability.\footnote{The equal predictive ability is immediate from the fact that the distribution of the random triple $(\textrm{Mix}_A, \textrm{Mix}_B, Y)$ is the same as the distribution of the random triple $(\textrm{Mix}_B, \textrm{Mix}_A, Y)$.}  However, individual replicates of Mix$_A$ and Mix$_B$ yield distinct scores.  The histogram at left in Figure~\ref{fig:DM} shows the $p$ values that arise from 400 replicates of Diebold--Mariano tests of Mix$_A$ (index $j$) versus Mix$_B$ (index $k$); as desired, the histogram is nearly uniform.  The histograms at middle and at right in Figure~\ref{fig:DM} show the $p$ values in the 400 tests of Mix$_A$ versus LM, and Mix$_B$ versus LM, respectively, which all lead to the rejection of the null hypothesis in favor of superior predictive ability of the LM model. 

\subsection{Comparison to the CSEP T-test}  \label{sec:T}

At this stage, let us compare to the CSEP T-test, which also aims to compare the predictive ability of two models.  As proposed by \citet[Section~2.3]{Rhoadesetal2011}, the CSEP T-test of equal predictive ability is based on quantities called the \textit{information gain}~($\mathrm{IG}_{(\hsp j,k)}$) and the \textit{information gain per earthquake}~($\mathrm{IG}_{(\hsp j,k)}^\mathrm{PE}$) of model $k$ over model $j$, respectively.  In Appendix~\ref{app:Poisson} we prove that 
\begin{equation}  \label{eq:IG}
\mathrm{IG}_{(\hsp j,k)} = T \left( \bar\myS^{(\hsp j)} - \bar\myS^{(k)} \right)
\qquad \textrm{and} \qquad
\mathrm{IG}_{(\hsp j,k)}^\mathrm{PE} = \frac{T}{N_T} \left( \bar\myS^{(\hsp j)} - \bar\myS^{(k)} \right) \! ,
\end{equation} 
where $N_T = \sum_{t = 1}^T \sum_{c = 1}^C y_{c,t}$, are both multiples of the Poisson score difference $\bar\myS^{(\hsp j)} - \bar\myS^{(k)}$.  Table~\ref{tab:IG} illustrates the relationships between the total Poisson score, the Poisson score difference, the information gain $\mathrm{IG}_{(\hsp j,k)}$, and the information gain per earthquake $\mathrm{IG}_{(\hsp j,k)}^\mathrm{PE}$ for our five forecast models, where the index $k$ stands for the LM model, the index $j$ for the model at hand, and where $T = 5514$ and $N_T = 1834$. 

\begin{table}[t]
\centering
\caption{Predictive performance according to the total Poisson score ($\bar\myS^{(\hsp j)})$, the Poisson score difference, information gain $\mathrm{IG}_{(\hsp j,k)}$, and information gain per earthquake ($\mathrm{IG}_{(\hsp j,k)}^\mathrm{PE}$) of the LM model ($k$) relative to the model at hand ($\hsp j$).  The best (lowest) score in each column is highlighted in green.  \label{tab:IG}}
\medskip
\begin{tabular}{lrrrr}
\toprule
& $\bar\myS^{(\hsp j)}$ & $\bar\myS^{(\hsp j)} - \bar\myS^{(k)}$ & $\mathrm{IG}_{(\hsp j,k)}$ & $\mathrm{IG}_{(\hsp j,k)}^\mathrm{PE}$ \\ 
\midrule
LM   &   2.71   &   0.000   &   0.000   &   0.000  \\
FCM  &   2.80   &   0.084   &   463.572   &   0.253  \\
LG   &   3.02   &   0.308   &   1697.038   &   0.925  \\
\midrule
SMA  &   2.73   &   0.020   &   111.393   &   0.061  \\
LRWA &   \green{2.69}   &   \green{$-0.018$}   &   \green{$-96.795$}   &   \green{$-0.053$} \\
\bottomrule
\end{tabular}
\end{table}

These relationships demonstrate two key insights.  First, $\mathrm{IG}_{(\hsp j,k)}$ and $\mathrm{IG}_{(\hsp j,k)}^\mathrm{PE}$ are perfectly valid tools for comparing the predictive performance of model forecasts in the form of expected earthquake counts.  Second, despite their derivation in terms of Poisson likelihoods in \citet{Rhoadesetal2011}, the use of $\mathrm{IG}_{(\hsp j,k)}$ and $\mathrm{IG}_{(\hsp j,k)}^\mathrm{PE}$ does not depend on the assumption of Poisson-distributed earthquake counts, nor does it rely on any other parametric distributions.  The only assumption that needs to be made is that the model forecasts represent expected counts in the technical sense of the expectation or mean of a random variable, as pointed out previously \citep{Brehmetal2021}.  Importantly, these arguments support the use of $\mathrm{IG}_{(\hsp j,k)}$ and $\mathrm{IG}_{(\hsp j,k)}^\mathrm{PE}$ in much broader settings than previously thought feasible. 

As noted before, additional procedures and assumptions are required to make formal statistical inference.  In this regard, the T-test proposed in Section~2.3 of \citet{Rhoadesetal2011} has shortcomings in the form of an incorrectly specified variance estimate.  As a consequence, the underlying test statistic has an incorrect standardization factor, and the $p$ value generated by the T-test generally fails to be uniform under the null hypothesis of equal predictive ability.  For details, we refer to Appendix~\ref{app:T}, where we demonstrate these issues in displays analogous to Table~\ref{tab:DM} and Figure~\ref{fig:DM}.  In this light, we recommend that the CSEP T-test be evolved into the Diebold--Mariano test based on spatially aggregated Poisson scores.  The two approaches rank models in the very same way, but they differ in the statistical properties of the tests; only the Diebold--Mariano test shows the desired standardization and, thus, the desired behavior under the null hypothesis.

We hasten to emphasize that the CESP T-test and the Diebold--Mariano test are closely related: They both are variants of the classical Student's $t$-test, and if one corrects for issues in the estimation of the variance as described in Appendix~\ref{app:T}, the CSEP T-test morphs into the Diebold--Mariano test.  Thus, we provide an improved version of the CSEP T-test by taking care of clerical issues that have not received sufficient attention thus far.

\section{Calibration, reliability, and discrimination}  \label{sec:calibration}

Up to now we have focused on comparative evaluation of the models --- we have asked whether a model is better than its competitors.  We now turn to the question of calibration and ask how well a model's forecasts agree with the outcomes.  Briefly, calibration is a joint property of the forecasts and the outcomes that summarizes the degree of mutual consistency.  It thus addresses the question whether a forecast provides reliable information about the outcomes.  This kind of testing has been originally named \textit{consistency tests}\/ since the first CSEP-related studies \citep[e.g.,][]{werner_retrospective_2010} and their predecessor \citep{Schoretal2007}; it puts the focus on formal statistical tests of the agreement between forecasts and observations given specific parametric assumptions.  In other scientific fields, these types of questions have been addressed via the \textit{calibration}\/ concept \citep[e.g.,][]{GBR2007, Wei2014, GneitResin2022}.  Similar to CSEP's consistency definition, calibration refers to the statistical consistency between the forecasts and the observations.  However, similar to the point process-oriented techniques proposed by \citet{Bray2013}, \citet{Thorarinsdottir2013}, and \citet{Bray2014}, our focus here is on calibration checks from a diagnostic perspective, where the goal is to identify, and eventually remedy, model deficiencies. 

\subsection{Conditional calibration}  \label{sec:conditionally_calibration}

Since our study deals with models that produce forecasts for the expected or mean number of earthquakes, we ask whether the forecasts are mean-calibrated, i.e., whether the forecasted means are consistent with the observed numbers of earthquakes.  To state this mathematically, we follow \citet{GneitResin2022} and conceptualize in terms of a random vector $(X,Y)$ where $X$ is the forecast and $Y$ the corresponding observation.  Then the forecast $X$ is \textit{conditionally mean-calibrated}, or simply \textit{calibrated}, if
\begin{equation}   \label{eq:mean_calibration}
\E ( Y \mid X) = X
\end{equation}
holds, i.e., if the conditional mean or expectation of the outcome, given any forecast value $X = x$, equals $x$.  In practice, mean-calibration implies, e.g., that, considering all grid cells and days on which an expected count of $x = 0.70$ earthquakes is forecasted, the number of observed earthquakes averages to 0.70.\footnote{In estimation problems, this type of property is commonly referred to as \textit{unbiasedness}.}  Hence, such a forecast is also termed \textit{reliable}.  Naturally, no real-world model can provide perfectly calibrated forecasts, thus we are more interested in how much it deviates from the ideal in~\eqref{eq:mean_calibration}.  The problem of a model's validation, which is a basic component of any scientific enterprise, requires further work, because it should also consider epistemic uncertainty \citep{Marzocchi2018}.  We intend to address this issue in future work.

\subsection{Mean-reliability curves}  \label{sec:reliability}

In situations where the forecasts take only a few different values, it is straightforward to check for mean-calibration by pooling forecasts with a specific value and, for each such group, comparing to the corresponding average value of the outcomes.  However, mean-forecasts usually take arbitrarily different values, such that this approach is inappropriate without cumbersome and possibly subjective binning of the forecasts.  The recently developed CORP (Consistent, Optimally binned, Reproducible, and PAV algorithm based) approach of \citet{Dimitetal2021} and \citet{GneitResin2022} addresses these issues and yields a mathematically principled, graphical assessment of mean-calibration.  While the initial development  by \citet{Dimitetal2021} concerned probability forecasts of binary outcomes, we follow \citet{GneitResin2022} in considering mean-forecasts of real-valued outcomes. 

Given a set of forecast--observation pairs $(x_1, y_1), \ldots, (x_n, y_n)$ the CORP approach uses nonparametric isotonic mean regression to calculate a sequence of (re) calibrated values $\hat{x}_1, \ldots , \hat{x}_n$, respectively.  The nonparametric isotonic regression approach is implemented by the classical pool adjacent violators (PAV) algorithm \citep{Ayer1955, PAVAinR}.  As described in Algorithm \ref{alg:PAV}, the PAV technique sorts the forecast--observation pairs and then partitions the index set $\{ 1, \ldots, n \}$ into bins $B_{k:l} = \{ k, \ldots, l \}$ of consecutive integers, which are pooled in an iterative fashion, until mean-calibration is achieved.  The resulting nonparametric isotonic regression curve yields a calibrated value $\hat{x}$ for the original forecast value $x$.  When viewed as a function of the original forecast values, and interpolated linearly in between, the calibrated values form a nondecreasing, piecewise linear curve that typically includes horizontal segments.

\begin{algorithm}[t]
\caption{PAV algorithm based on forecast--outcome tuples $(x_1, y_1)$, \ldots, $(x_n, y_n)$.  The rearrangement of the indices allows for a succinct description of the algorithm, which is adopted from \citet[p.~3249]{GneitResin2022}.  \label{alg:PAV}} 
\SetAlgoLined
\KwIn{tuples $(x_1,y_1), \ldots, (x_n,y_n)$}
\KwOut{calibrated values $\widehat{x}_1, \ldots, \widehat{x}_n$}
\smallskip
rearrange indices such that $x_1 \leq \cdots \leq x_n$ \\
partition into bins $B_{1:1}, \ldots, B_{n:n}$ and let $\widehat{x}_i = y_i$ for $i = 1, \ldots, n$ \\
\While{there are bins $B_{k:i}$ and $B_{(i+1):l}$ such that
  $\widehat{x}_1 \leq \cdots \leq \widehat{x}_i$ and $\widehat{x}_i > \widehat{x}_{i+1}$}
{merge $B_{k:i}$ and $B_{(i+1):l}$ into $B_{k:l}$ and let $\widehat{x}_i = (y_k + \cdots + y_l)/(l - k + 1)$ for $i = k, \ldots, l$}
\end{algorithm}
 
The calibrated value $\hat{x}$ is an estimate of the conditional expectation $\E ( Y \mid X = x)$ and thus it should be close to $x$ under the mean-calibration property in \eqref{eq:mean_calibration}.  The \textit{mean-reliability curve}\/ is the graph of the piecewise linear function that arises from an interpolation of the points $(x_1, \hat{x}_1), \ldots, (x_n, \hat{x}_n)$, where we assume, for simplicity of the description, that $x_1 \leq \cdots \leq x_n$.  Horizontal segments in a mean-reliability curve then correspond to distinct forecast values $x_i \leq \cdots \leq x_j$ such that $\hat{x}_i = \cdots = \hat{x}_j$.  A perfectly calibrated forecast has a mean-reliability curve directly on the diagonal.  Major deviations from the diagonal in the mean-reliability diagram indicate a lack of mean-calibration and can be interpreted diagnostically.  

\begin{figure}[tbp]
\centering
\includegraphics[scale=1.0]{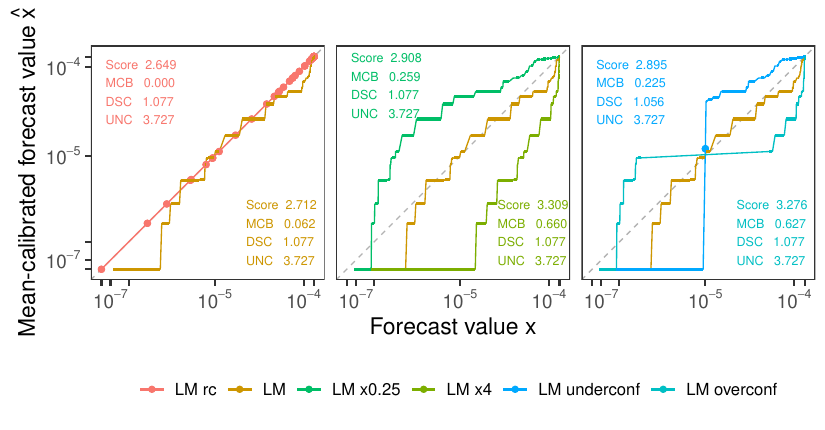}
\vspace{-3mm}
\caption{Mean-reliability curves for the LM forecast and artificially manipulated variants thereof as described in the text, using an empirical CDF transform aggregated over all models to scale the axes: original and (re) calibrated forecast (left); over- and underforecast (middle); over- and underconfident forecast (right).  The tickmarks correspond to non-transformed, original values of $0$, $10^{-7}$, $10^{-6}$, $10^{-5}$, $10^{-4}$, and $0.88$, respectively.  We also show the associated total Poisson score at \eqref{eq:Sbar} and its decomposition into miscalibration (MCB), discrimination (DSC), and uncertainty (UNC) components from \eqref{eq:decomposition}, as described in Section~\ref{sec:MCB_DSC}.  \label{fig:reliability_simulation}}
\end{figure}

In the spatio-temporal setting of earthquake forecasting, we have $n = C \times T$ and the forecast--observation pairs $(x_1, y_1), \ldots, (x_n, y_n)$ correspond to the collection of tuples  
\[
\left( x_{c,t}^{(\hsp j)}, y_{c,t} \right)_{c = 1, \ldots, C; \; t = 1, \ldots, T}
\]
with forecast values $x_{c,t}^{(\hsp j)}$ from model $j$ and corresponding counts $y_{c,t}$ of target earthquakes.  For an initial illustration, Figure~\ref{fig:reliability_simulation} provides mean-reliability diagrams for the LM forecast and artificially manipulated variants thereof.  As the forecast values are mostly very small, the typical linear--linear diagram is hard to interpret in this setting.  Instead, we plot reliability diagrams on an empirical CDF scale on both the horizontal and the vertical axes.  Specifically, we use the empirical CDF of the forecast values over all five models (LM, FCM, LG, SMA, and LRWA), all grid cells, and all test times to scale the axes.  For example, a linear value of 0.20 indicates the 20th percentile in this aggregated distribution, and the tickmarks correspond to $0$, $10^{-7}$, $10^{-6}$, $10^{-5}$, $10^{-4}$, and the maximum $0.88$ of the forecast values, respectively.  The mean-reliability curve for the LM model in the left panel is rather close to the diagonal, as desired, though it remains below the diagonal, and in fact at zero, for forecast values up to about $3.3 \cdot 10^{-6}$.  The same panel also shows the mean-reliability curve for the perfectly (re) calibrated LM forecast (LM rc), which is directly on the diagonal, as enforced by the PAV algorithm. 

The middle panel in Figure~\ref{fig:reliability_simulation} shows mean-reliability curves that result from underforecasting and overforecasting, respectively.  Specifically, the LM $\times 4$ forecast multiplies the expected count under the LM model by a factor of 4.  This results in overforecasts, with observed earthquake counts that are substantially lower on average than the manipulated forecast value and, therefore, a mean-reliability curve that is well below the diagonal.  The LM $\times 0.25$ forecast divides the LM forecast by a factor of 4, which results in underforecasts and a mean-reliability curve mostly well above the diagonal.  The right panel shows an overconfident variant of the LM model (`LM overconf') that issues forecast values too far out in the tails, and an underconfident version (`LM underconf') that pushes the forecast values away from the extreme tails.\footnote{Specifically, the LM overconf forecast agrees with LM $\times 0.25$ for an LM model output $\leq 10^{-5}$ and agrees with LM $\times 4$ for an LM model output $> 10^{-5}$.  The LM underconf forecast agrees with LM $\times 4$ for an LM model output $\leq 0.25 \cdot 10^{-5}$, agrees with LM $\times 0.25$ for an LM model output $\geq 4 \cdot 10^{-5}$, and equals $10^{-5}$ otherwise.}  The mean-reliability curves show the characteristic S-shape for an underconfident forecast, and the typical inverse S-shape for an overconfident forecast, respectively.

\begin{figure}[tbp]
\centering
\includegraphics[scale=1.0]{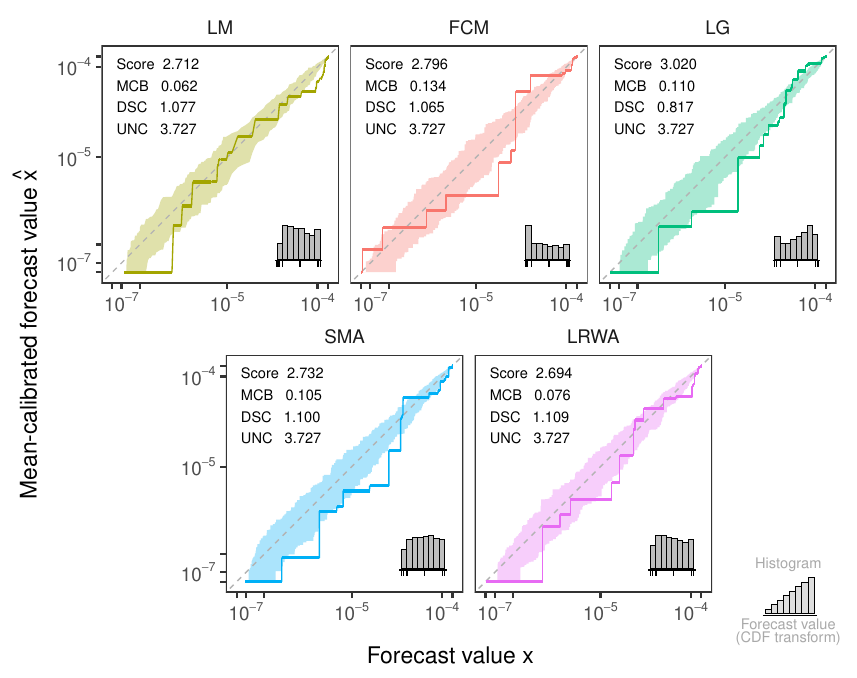}
\caption{Mean-reliability curves as in Figure~\ref{fig:reliability_simulation}, but for the five forecast models, along with 90 per cent consistency bands (shaded), using an empirical CDF transform aggregated over all models to scale the axes.  The inset histograms depict the distribution of a model's forecast values after this transform.  \label{fig:reliability_CSEP}}
\end{figure}

Figure~\ref{fig:reliability_CSEP} shows reliability diagrams for the five forecast models.  In practice, mean-reliability curves deviate from the diagonal even when the hypothesis of mean-calibration is true, for reasons of sampling variability alone.  We use \textit{consistency bands}\/ to address the question whether an observed deviation from the diagonal can reasonably be attributed to random (sampling) fluctuations alone, despite the forecasts being calibrated.  Thus, the 90 per cent consistency bands in Figure~\ref{fig:reliability_CSEP} range (pointwise) from the 5th to the 95th percentile of mean-reliability curves that are sampled under the hypothesis of mean-calibration, using the method described in Appendix~\ref{app:consistency_bands}.  In addition to the mean-reliability curves, we show histograms for the forecast values for the model at hand, using the same empirical CDF scale based on all five models aggregated, as described above and used for the mean-reliability curves.  The LM model is well calibrated as its mean-reliability curve fluctuates around the diagonal mostly within the 90 per cent consistency band that quantifies the variation of the calibration curve under the hypothesis of mean-calibration.  In contrast, the calibration curves of the FCM, LG, and SMA models oftentimes leave the consistency bands.  Apart from the FCM model, which frequently issues very small forecasts, as indicated by the spike at the lower range of the inset histogram, small and moderate forecast values generally tend to be too large (i.e., overforecasting) since the calibration curve is beneath the diagonal for small forecasted means, and large forecast values (on the CDF scale) tend to be too small (i.e., underforecasting), for a slight indication of underconfidence.

\subsection{Miscalibration--discrimination (MCB--DSC) diagrams}  \label{sec:MCB_DSC}

As illustrated in Figures~\ref{fig:reliability_simulation} and \ref{fig:reliability_CSEP}, mean-reliability curves allow for a diagnostic interpretation of a model's performance in terms of calibration.  In a related development, and also based on the CORP (re) calibration method, \citet{GneitResin2022} introduce a decomposition of the average score 
\begin{equation}  \label{eq:Sbar_1_n}
\bar\myS = \frac{1}{n} \sum_{i = 1}^n \myS(x_i, y_i)
\end{equation} 
for a collection $(x_1, y_1), \ldots, (x_n, y_n)$ of forecast--observation pairs that allows for an interpretation in terms of (mis) calibration and discrimination ability.  Specifically, let $\hat{x}_1, \ldots, \hat{x}_n$ denote the PAV (re) calibrated forecasts that correspond to $x_1, \ldots, x_n$, respectively, and let $\hat{x}_\mathrm{mg} = \frac{1}{n} \sum_{i = 1}^{n} y_i$ denote the overall (or marginal) mean, which serves as a simple baseline mean-forecast.  Given a consistent scoring function $\myS$ we let
\[
\bar\myS_\mathrm{rc} = \frac{1}{n} \sum_{i = 1}^n \myS(\hat{x}_i, y_i) \qquad \textrm{and} \qquad 
\bar\myS_\mathrm{mg} = \frac{1}{n} \sum_{i = 1}^n \myS(\hat{x}_\mathrm{mg}, y_i)
\]
denote the average score of the (re) calibrated forecast and the marginal forecast, respectively.  We can then define the \textit{miscalibration}~(MCB), \textit{discrimination}~(DSC), and \textit{uncertainty}~(UNC) components of the average score $\bar\myS$ as
\begin{equation}  \label{eq:score_decomp}
\mathrm{MCB} = \bar\myS - \bar\myS_\mathrm{rc}, \qquad
\mathrm{DSC} = \bar\myS_\mathrm{mg} - \bar\myS_\mathrm{rc}, \qquad \textrm{and} \qquad 
\mathrm{UNC} = \bar\myS_\mathrm{mg},
\end{equation}
respectively.  The MCB term compares the original forecast to the calibrated one, and attains its minimal value of zero if, and only if, the original forecast is calibrated; the DSC term compares the best constant forecast to the calibrated forecast, and attains its minimal value of zero if, and only if, $x_i = x_1$ for $i = 1, \ldots, n$ \citep{GneitResin2022}.  In other words, the MCB reflects how unreliable a model is, whereas the DSC  term reflects how well a model can sort out different scenarios (e.\,g., here especially background seismicity and triggered seismicity).  The UNC term is independent of the issued forecast values and quantifies the variability of the outcomes.  All three components are nonnegative and yield the CORP score decomposition
\begin{equation}  \label{eq:decomposition}
\bar\myS = \mathrm{MCB} - \mathrm{DSC} + \mathrm{UNC}. 
\end{equation}
Evidently, low MCB and high DSC terms are desirable, and a low average score $\bar\myS$ might stem from low MCB, or high DSC, or both.  Analogously, a high average score might stem from high MCB, or low DSC, or both.

For an illustration of the decomposition \eqref{eq:decomposition} under the Poisson scoring function on manipulated LM forecasts we return to Figure~\ref{fig:reliability_simulation}, where $n = C \times T$.  We note that the (re) calibrated LM forecast shares DSC with the original LM model, but has lower (namely, zero) MCB.  The LM $\times 4$ and LM $\times 0.25$ forecasts also share DSC with the original LM model, but have much larger MCB.  These observations highlight that DSC only reflects the relative context of a model's forecast values.  The `LM overconf' and `LM underconf' forecasts degrade MCB and/or DSC.  Similarly, Figure~\ref{fig:reliability_CSEP} illustrates the decomposition \eqref{eq:decomposition} for the five forecast models under the Poisson scoring function.  

\begin{figure}[tb]
\centering
\includegraphics[scale=1.0]{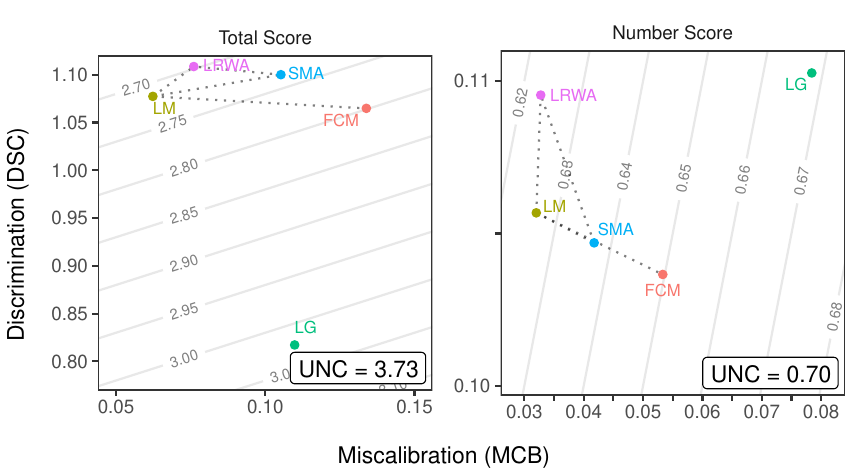}
\caption{MCB--DSC diagram for the five forecast models under the Poisson scoring function, for (Left) the total score at \eqref{eq:Sbar} and (Right) the number score at \eqref{eq:Sbar_number}.  For the pairs of models connected by (dotted) lines, the (two-sided) Diebold--Mariano test does not reject the null hypothesis of equal predictive ability in terms of the Poisson scoring function at the 0.10 significance level.  \label{fig:MCB_DSC_CSEP}}
\end{figure}

For a more succinct presentation of the CORP score decomposition at \eqref{eq:decomposition} we consider miscalibration--discrimination (MCB--DSC) diagrams, as introduced in related contexts by \citet{Dimitriadis2024} and \citet{Gneiting2023}.  An MCB--DSC diagram plots, for each model considered, DSC on the vertical axis versus MCB on the horizontal axis.  In view of the relationship at \eqref{eq:decomposition}, models with an equal average score $\bar\myS$ are represented on parallel lines with positive slope,\footnote{As the relationships in \eqref{eq:IG} demonstrate, these lines also indicate equal information gain, or equal information gain per earthquake.} and models at upper left in the diagram score best, with low MCB and high DSC, whereas models at lower right score worst, with high MCB and low DSC.  Models that are located (roughly) on one of the parallel lines have a similar average score but trade MCB for DSC, or vice versa.  The left panel in Figure~\ref{fig:MCB_DSC_CSEP} shows an MCB--DSC diagram for the five forecast models from Figure~\ref{fig:reliability_CSEP} using the total score \eqref{eq:Sbar}.  Accordingly, the inferior performance of the LG model stems from a substantial lack of discrimination ability (evidently from Figure~\ref{fig:score.IG.IGPE}, it performs well for quiescent periods, i.e., background seismicity, but worst during the sequences in central Italy 2009 and 2016).  The LRWA and LM models outperform their competitors and show pleasant balance between calibration and discrimination.  The differences in the total score between the LRWA, LM, SMA, and FCM models are modest.

The right panel in Figure~\ref{fig:MCB_DSC_CSEP} shows an MCB--DSC diagram for spatially aggregated model forecasts, as shown in the top panel of Figure~\ref{fig:forecasts}.  Specifically, the \textit{number score}\/ of model $j$, 
\begin{equation}  \label{eq:Sbar_number} 
\bar\myS_\#^{(\hsp j)} = \frac{1}{T} \sum_{t = 1}^T \myS \! \left( \sum_{c = 1}^C x_{c,t}^{(\hsp j)}, \sum_{c = 1}^C y_{c,t} \right) \! ,
\end{equation}
is the average score of the spatially aggregated expected numbers of target earthquakes.\footnote{Interestingly, if there are no target earthquakes at a given time $t$, then the models' contributions to the total score equal their contributions to the number score.  This follows from the fact that if $\myS$ is the Poisson scoring function at \eqref{eq:Poisson} then
\[
\bar\myS_t^{(\hsp j)} 
= \sum_{c = 1}^C \myS(x_{c,t}^{(\hsp j)}, y_{c,t}) 
= \myS \! \left( \sum_{c = 1}^C x_{c,t}^{(\hsp j)}, \sum_{c = 1}^C y_{c,t} \right)
+ \sum_{c = 1}^C \myS \! \left( \frac{x_{c,t}^{(\hsp j)}}{\sum_{b = 1}^C x_{b,t}^{(\hsp j)}}, y_{c,t} \right) - 1. 
\]
The middle term on the right-hand side reduces to the constant 1 if $y_{c,t} = 0$ for $c = 1, \ldots, C$, which proves the claim.}  In typical practice, analyses based on total scores as in \eqref{eq:Sbar} and analyses based on number scores as in \eqref{eq:Sbar_number} yield similar, if not identical, model rankings.  Yet, we prefer total scores since comparisons of spatially aggregated forecasts in number scores ignore regional differences in predictive performance.  For example, a model that overforecasts in some part of the testing region and underforecasts in another part gets penalized by the total score in \eqref{eq:Sbar}, but if the effects balance each other, the model's deficiencies may not be discernible from the number score in \eqref{eq:Sbar_number}.  In the spatially aggregated setting, the number of forecast cases dwindles to $n = T$, as opposed to $n = C \times T$ for the total score at the grid cell level.  While the rankings of the five forecast models under the total score and under the number score are identical, the LG model now lacks predominantly in terms of MCB.  

As noted, forecasts in the form of expected counts allow for aggregation at any desired spatial and/or temporal level, and we encourage the development of multi-resolution evaluation approaches in the spirit of \citet{Asim2023}.

\subsection{Comparison to existing CSEP methodology}  \label{sec:comparison_calibration}

Our focus in this section has been on diagnostics, where the goal is to identify, and eventually remedy, model deficiencies in terms of the notions of (mis) calibration, also referred to as reliability, and discrimination ability.  In contrast, extant CSEP methodology for model evaluation has put the focus on formal statistical tests of agreement between forecasts and outcomes, frequently under the label of \textit{earthquake likelihood model testing}\/ \citep{KaganJack1995, Schoretal2007}.  For example, the CSEP L-test, CL-test, and N-test operate under the (questionable) joint hypothesis that the modeled expected cell counts are correct, and that each observed count has a Poisson distribution.  A model then might fail the test when either one or both of the hypotheses are false.  In particular, this happens when the expected cell counts are perfectly correct, but the observed counts do not come from the Poisson distribution.  Conversely, while a passed test expresses the fact that there is no evidence against either of the hypotheses, a model might pass the test, and yet be inferior to competitors that fail the test.  Thus, the outcomes of these tests may be misleading, except for cases where the model issues full-distribution forecasts that are explicitly Poisson distributions.  

The Poisson assumption was avoided by \citet{savran2020}, who tested full-distribu\-tion (i.e., catalog-based) earthquake forecasts with nonparametric analogs to the grid-based CSEP tests (N-, M-, L-, and S-test).  These tests were implemented in (py)CSEP \citep{savran2022pycsep}, but they only address consistency/calibration of full-distribution forecasts, not comparative evaluations.  The tools proposed in our paper are also nonparametric and do not rely on a Poisson assumption.  They are straightforward to implement and, in addition to model ranking via scoring functions, they can be used diagnostically, via mean-reliability curves and MCB--DSC diagrams, to identify, and eventually remedy, model deficiencies.  In addition, the proposed methods facilitate comparing also full-distribution forecasts. 

Worthy of note, as for the current CSEP testing implementation, both calibration and consistency tests do not account for the so-called epistemic uncertainty, considering only one model or the (weighted) mean of a set of models.  The consequences of that have been discussed in depth by \citet{Marzocchi2014}.  Here we just mention that a proper model validation can be made only considering such an uncertainty.  Hence, model consistency and calibration have mostly an heuristic value that tells us how much and how the observations deviate from the forecasts.

\section{Conclusion}  \label{sec:discussion}

In this study we considered short-term earthquake forecasting models for the Italian CSEP testing region and evaluated their forecasts (in the form of expected earthquake counts) with consistent scoring functions, mean-reliability diagrams, and miscalibration--discrimination (MCB--DSC) diagrams.  We illustrated that evaluation methods based on scoring functions offer a comprehensive assessment of the models.  They yield a broader perspective on the existing CSEP testing framework and provide a valuable tool box for earthquake model assessment.  Specifically, we make the following recommendations. 
\begin{itemize}   
\item There are many consistent scoring functions for forecasts in the format of expected counts, with the Poisson score playing a distinguished role in low probability environments.  Use Murphy diagrams to consider all consistent scoring functions simultaneously; the area under a logarithmic Murphy curve equals the average Poisson score. 
\item Retain the established practice of binary model comparisons via the information gain per earthquake \citep[$\mathrm{IG}^\mathrm{PE}$:][]{Ve-Jo2003, HarteVe-Jo2005, Rhoadesetal2011} measure.  The $\mathrm{IG}^\mathrm{PE}$ measure equals (up to a factor) the difference in the average Poisson score (Eq.~\ref{eq:IG}) and is much broader applicable than previously thought, as its use does not depend on any parametric assumptions. 
\item If formal inference is the goal, which depends on accurate uncertainty quantification, replace the CSEP T-test by its close relative, the Diebold--Mariano test based on spatially aggregated scores, which comprehensively accounts for sampling variability.
\item Use mean-reliability curves as a diagnostic tool to assess calibration. 
\item An average score $\bar\myS$ can be decomposed into MCB, DSC, and UNC components.  Compare and investigate them for multiple models in MCB--DSC diagrams. 
\end{itemize}
Importantly, the evaluation methods proposed here are model-agnostic, i.e., they can be applied to any type of forecast (in the form of expected earthquake counts), regardless of the model or technique that produced them.  Forecast and data characteristics such as spatial-temporal clustering and heavy tailed distributions affect the intrinsic variability in these methods, which gets quantified via the denominator in the Diebold--Mariano statistic \eqref{eq:z} and consistency bands in mean-reliability diagrams.  While further work in these directions may be the focus of future research, we are confident that these tools will help pave the way to improvements in the evaluation, selection, and development of earthquake forecasting models.

It is important to stress that our results hold only in terms of mean-forecasts.  In particular, we do not claim that the LRWA and LM models are superior in all facets.  The results indicate that they are slightly better at producing mean-forecasts for the number of earthquakes.  A key reason for this is that the forecasts of the LM model are well-calibrated and overall higher than the forecasts of its competitors.  Hence, the LM model is not as severely underforecasting as the LG and FCM models on days when (many) earthquakes happen. 

As emphasized by \citet{Nandanetal2019} it would be best if these forecasts specified the full distribution of earthquake counts, as this provides most information on the likely implications of different actions.  Yet, full-distribution forecasts may not be available due to their complexity, reporting traditions, ease of forecast communication, or the need for aggregation.  However, full-distribution forecasts can be supplied in the form of a collection of synthetic catalogs (also called stochastic event sets) of future earthquakes.\footnote{In their experiments, \citet{Nandanetal2019} use a collection of five million synthetic catalogs.}  Importantly, if collections of synthetic catalogs are interpreted as full-distribution forecasts, we can deduce mutually consistent full (empirical) distribution forecasts at any desired level of spatial and/or temporal aggregation.  The evaluation methods proposed in this paper also apply to full-distribution forecasts, by converting them to the implied mean-forecasts, as implemented here, which does not require smoothing nor parametric assumptions.

\section*{Data Availability}

The tables and figures were produced with R \citep{R}.  Code for reproduction is available on \url{https://github.com/jbrehmer42/Earthquakes_Italy}, where we also provide details on slight differences to previously reported scores (see footnote 4 describing Table 1).  Data are available from Marcus Herrmann (\texttt{marcus.herrmann@unina.it}) upon request.

\section*{Acknowledgments}

Jonas Brehmer, Tilmann Gneiting, and Kristof Kraus are grateful for support by the Klaus Tschira Foundation.  Marcus Herrmann and Warner Marzocchi were supported by the ‘Multi-Risk sciEnce for resilienT commUnities undeR a changiNg climate’ (RETURN) project, funded by the European Union's NextGenerationEU and the Italian Ministry of University and Research (MUR) under the National Recovery and Resilience Plan (PNRR; Code PE0000005).  We thank two anonymous reviewers, Johannes Bracher, Alexander Jordan, and Johannes Resin for helpful discussions and remarks.

\bibliographystyle{style1}
\bibliography{literature}

\appendix

\renewcommand\thesection{\Alph{section}}
\setcounter{section}{0}

\renewcommand\thefigure{\thesection\arabic{figure}}
\counterwithin{figure}{section}

\renewcommand\thetable{\thesection\arabic{table}}
\counterwithin{table}{section}

\renewcommand\theequation{\thesection\arabic{equation}}
\counterwithin{equation}{section}

\renewcommand{\textfraction}{0.0}       

\section*{Appendices}

\section{Consistent scoring functions}  \label{app:consistent}

\subsection{Technical details for the Bregman representation}  \label{app:Bregman}

Consider the Bregman representation \eqref{eq:Bregman}, namely, 
\[
\myS(x,y) = \phi(y) - \phi(x) - \phi'(x)(y - x), 
\]
for a scoring function that is consistent for the mean-functional.  For simplicity we ignore the function $h(y)$, and we recall that $\phi$ is a convex function with subgradient $\phi'$.  Regarding our setting of count data, where the mean-forecast $x \geq 0$ is a nonnegative number, whereas the outcome $y \in \{ 0, 1, \ldots \}$ is a nonnegative integer, we distinguish three cases for this representation. 

In case I the convex function $\phi$ is continuous and differentiable on the nonnegative halfaxis with $\phi(0)$ and $\phi'(0)$ being real numbers.  A key example is $\phi(x) = x^b$ for $b > 1$.  Then the score $\myS(x,y)$ is well-defined and finite for all $x \geq 0$ and $y \in \{ 0, 1, \ldots \}$.

In case II the convex function $\phi$ is defined on the strictly positive halfaxis with $\phi(0) = \lim_{x \downarrow 0} \phi(x)$ being finite, but $\phi'(0) = \lim_{x \downarrow 0} \phi'(x) = - \infty$.  A key example is $\phi(x) = - x^b$ for $0 < b < 1$.  Then the score $\myS(x,y)$ is well-defined and finite for $x > 0$ and $y \in \{ 0, 1, \ldots \}$.  To allow for the mean-forecast $x = 0$, we define $\myS(0,0) = 0$ and $\myS(0,y) = + \infty$ for $y \in \{ 1, 2, \ldots \}$, which retains consistency. 

In case III the convex function $\phi$ is defined on the strictly positive halfaxis with $\phi(0) = \lim_{x \downarrow 0} \phi(x) = + \infty$ and $\phi'(0) = \lim_{x \downarrow 0} \phi'(x) = - \infty$.  A key example is $\phi(x) = x^b$ for $b < 0$.  Then the score $\myS(x,y)$ is well-defined and finite for $x > 0$ and $y \in \{ 1, 2, \ldots \}$, but there is no obvious extension that maintains consistency. 

\subsection{Extended Patton family of consistent scoring functions}  \label{app:Patton}

The original \textit{Patton family}\/  of consistent scoring functions for an expected value \citep{Patton2011} is parameterized in terms of a real-valued index $b$ and defined via
\begin{equation}  \label{eq:Patton}
\myS_b(x,y) = 
\left\lbrace
\begin{array}{ll}
\dfrac{y^b - x^b}{b (b-1)} - \dfrac{x^{b-1}}{b-1} (y - x) ,  & b \notin \{ 0, 1 \}, \\
\dfrac{y}{x} - \log \dfrac{y}{x} - 1,                        & b = 0, \rule{0mm}{7mm} \\
y \log \dfrac{y}{x} - (y - x),                               & b = 1, \rule{0mm}{7mm}
\end{array}
\right.
\end{equation}
for an expected value $x > 0$ and an outcome $y > 0$.  The members $\myS_b$ of the Patton family are strictly consistent scoring functions for an expected value, as they are of the form~\eqref{eq:Bregman} with the strictly convex function
\begin{equation}   \label{eq:phi_Patton}
\phi_b(x) =
\left\lbrace
\begin{array}{ll}
\dfrac{x^b}{b (b-1)},  & b \notin \{ 0, 1 \}, \\
- \log x,              & b = 0, \rule{0mm}{5mm} \\
x \log x,              & b = 1, \rule{0mm}{5mm} 
\end{array}
\right.
\end{equation}
and a certain choice of $h(y)$.  The Patton family nests both the Poisson scoring function ($b = 1$) at \eqref{eq:Poisson} and the quadratic scoring function ($b = 2$) at \eqref{eq:quadratic}, up to a constant factor and terms that only depend on $y$.  However, the specific choice of $h(y)$ yields expressions for $\myS_b(x,y)$ that can be undefined when $x = 0$ or $y = 0$, and so the Patton functions need to be modified to evaluate forecasts of earthquake counts. 

To accommodate count data, while retaining the nesting property, we define the \textit{extended Patton family}\/  via
\begin{equation}  \label{eq:extended_Patton}
\myS_b^0(x,y) = \myS_b(x,y) - \myS_b(1, y) + \frac{1}{2} y^b - \frac{b}{2} y + \frac{3-b}{2}
\end{equation}
for $x > 0$ and $y > 0$.  The members $\myS_b^0$ of the extended Patton family are still strictly consistent, since $\myS_b^0(x,y)$ agrees with $\myS_b(x,y)$ up to terms that depend on $y$ only, and thus they are of the form \eqref{eq:Bregman} with the same choice of $\phi_b$.

To extend $\myS_b^0(x,y)$ to $x \geq 0$ and $y \in \{ 0, 1, \ldots \}$ we proceed as discussed in Appendix~\ref{app:Bregman}.  If $b > 1$ we are in case I and extend by continuity.  If $0 < b \leq 1$ we are in case II and extend by setting $\myS_b^0(0,0) = 0$ and $\myS_b^0(0,y) = + \infty$ for $y \in \{ 1, 2, \ldots \}$.  If $b \leq 0$ we are in case III, when there is no obvious extension.  The expression in \eqref{eq:extended_Patton} yields $\myS_b^0(x,y) = x - y \log x$ for $x > 0$ and $y \in\{ 0, 1, \ldots \}$ when $b = 1$, and $\myS_b^0(x,y) = \frac{1}{2} (y - x)^2$ for $x \geq 0$ and $y \in\{ 0, 1, \ldots \}$ when $b = 2$, and thus the extended Patton family nests both the Poisson scoring function and ($\frac{1}{2}$ times) the quadratic scoring function.

\section{Further technical details and results for Section \ref{sec:scoring}}  \label{app:technical}

In this appendix, we provide further technical details and supplementary results for Section~\ref{sec:scoring}.  Throughout, the scoring function $\myS$ is the Poisson scoring function at \eqref{eq:Poisson}.  Whenever feasible, we present formulas in the generic spatio-temporal setting of earthquake forecasts, where there are $C$ grid cells and $T$ regularly spaced test times, with earthquake count $y_{c,t}$ in grid cell $c \in \{ 1, \ldots, C \}$ at time $t \in \{ 1, \ldots, T \}$.  In the specific case of the OEF-Italy there are $C = 8993$ grid cells and $T = 5514$ test times. 

\subsection{Quantities in Figure~\ref{fig:score.IG.IGPE} panels}  \label{app:figure}

We give technical details for the quantities displayed in the four panels of Figure~\ref{fig:score.IG.IGPE}.  To this end, suppose that model $j$ issues forecasts in the form of expected cell counts $x_{c,t}^{(\hsp j)}$.  As defined at \eqref{eq:Sbar_t}, the spatially aggregated score of model $j$ at time $t$ is 
\begin{equation}  \label{appeq:Sbar_t}
\bar\myS_t^{(\hsp j)} = \sum_{c = 1}^C \myS \! \left( x_{c,t}^{(\hsp j)} , y_{c,t} \right),  
\end{equation}
which is the quantity plotted against time $t$ in the first (top) panel in Figure~\ref{fig:score.IG.IGPE}, where the index $j$ stands for the LM, FCM, LG, SMA, and LRWA model, respectively.  The second panel in Figure~\ref{fig:score.IG.IGPE} shows the daily score difference 
\begin{equation}  \label{appeq:Sbar_t_diff}
d_t^{\hsp (\hsp j,k)} = \bar\myS_t^{(\hsp j)} - \bar\myS_t^{(k)}, 
\end{equation}
where the index $j$ stands for the FCM, LG, SMA, and LRWA model, and the index $k$ for the LM model.  

Not surprisingly, plots of daily Poisson scores or daily Poisson score difference are highly irregular.  To obtain smoother curves, one can plot the cumulative score difference
\begin{equation}  \label{appeq:cum_diff}
D_t^{\hsp (\hsp j,k)} = \sum_{u = 1}^t d_u^{\hsp (\hsp j,k)}
\end{equation}
against time $t$, as in Figure~2 of \citet{Taronietal2018} and in the third panel of our Figure~\ref{fig:score.IG.IGPE}.  Alternatively, Figures~5, 6, and C2 of \citet{HM2023} and the fourth panel of our Figure~\ref{fig:score.IG.IGPE} plot the normalized quantity 
\begin{equation}  \label{appeq:cum_diff_n}
D_t^{\hsp (\hsp j,k)} / N_t, 
\end{equation}
where
\begin{equation}  \label{appeq:N}
N_t = \sum_{u = 1}^t \sum_{c = 1}^C y_{c,u}
\end{equation}
is the cumulative count of target earthquakes over the test region and test times $u \in \{ 1, \ldots, t \}$.  We note that the quantities \eqref{appeq:Sbar_t} to \eqref{appeq:cum_diff_n}, which are displayed in the four rows of Figure~\ref{fig:score.IG.IGPE}, are all negatively oriented, i.e., the smaller the better.  

\subsection{Total Poisson score, information gain, and information gain per earthquake in spatio-temporal settings}  \label{app:Poisson}

Let us recall from \eqref{eq:Sbar} that the total score of model $j$ is 
\begin{equation*}
\bar\myS^{(\hsp j)} = \frac{1}{T} \sum_{t = 1}^T \bar\myS_t^{(\hsp j)}, 
\end{equation*}
where $\bar\myS_t^{(\hsp j)}$ is the spatially aggregated score for model $j$ on day $t$ from \eqref{appeq:Sbar_t}.  The $z$ statistic \eqref{eq:z} of the Diebold--Mariano test of equal predictive ability of model $j$ and model $k$ is a multiple of the difference 
\begin{equation}  \label{appeq:Sbar_diff}
\bar\myS^{(\hsp j)} - \bar\myS^{(k)}, 
\end{equation}
where $\myS$ is the Poisson score.  We proceed to prove the claims in \eqref{eq:IG} that the information gain $\mathrm{IG}_{(\hsp j,k)}$ and the information gain per earthquake $\mathrm{IG}_{(\hsp j,k)}^\mathrm{PE}$ of model $k$ over model $j$ also are multiples of the Poisson score difference at \eqref{appeq:Sbar_diff}. 

We start with time $t$ fixed and consider the definition of the information gain in the purely spatial setting of Eq.~(17) in \citet[p.~740]{Rhoadesetal2011}.  In this simpler setting, the information gain of model $k$ over model $j$ at time $t$ is   
\begin{equation}  \label{appeq:IG_t}
\sum_{c = 1}^C y_{c,t} \left( \log x_{c,t}^{(k)} - \log x_{c,t}^{(\hsp j)} \right) - \sum_{c = 1}^C \left( x_{c,t}^{(k)} - x_{c,t}^{(\hsp j)} \right) \! .  
\end{equation}
To see the equivalence of our Eq.~\eqref{appeq:IG_t} and Eq.~(17) in \citet{Rhoadesetal2011}, note the specification of the quantity $i_k$ at the top of their page 731 and the fact that 
\[
\sum_{c = 1}^C y_{c,t} \left( \log x_{c,t}^{(k)} - \log x_{c,t}^{(\hsp j)} \right) 
= \sum_{z=1}^\infty \; z \sum_{c = 1}^C \one{y_{c,t} = z} \left( \log x_{c,t}^{(k)} - \log x_{c,t}^{(\hsp j)} \right)
\]
represents the score differences in bins with observed target earthquakes.  Summing the expression for the information gain at time $t$ in \eqref{appeq:IG_t} over testing times $t \in \{ 1, \ldots, T \}$, the spatio-temporal information gain $\mathrm{IG}_{(\hsp j,k)}$ of model $k$ over model $j$ is 
\begin{equation}  \label{appeq:IG}
\mathrm{IG}_{(\hsp j,k)} = \sum_{t = 1}^T \sum_{c = 1}^C y_{c,t} \left( \log x_{c,t}^{(\hsp j)} - \log x_{c,t}^{(k)} \right) 
- \sum_{t = 1}^T \sum_{c = 1}^C \left( x_{c,t}^{(\hsp j)} - x_{c,t}^{(k)} \right) \! . 
\end{equation}
Invoking Eqs.~\eqref{eq:Poisson}, \eqref{eq:Sbar_t}, \eqref{eq:Sbar}, and \eqref{appeq:Sbar_diff}, we see that the information gain $\mathrm{IG}_{(\hsp j,k)}$ is of the form stated in Eq.~\eqref{eq:IG}.  

The expression for the information gain per earthquake $\mathrm{IG}_{(\hsp j,k)}^\mathrm{PE} = \mathrm{IG}_{(\hsp j,k)} / N_T$ in Eq.~\eqref{eq:IG} is now immediate.  We note that in the case of OEF-Italy, due to the overlap of the seven-day forecast periods, the earthquake count $N_T$ equals seven times the number of unique target earthquakes in the catalog.

\subsection{Results for the CSEP T-test}  \label{app:T}

As we have demonstrated in the previous section, the CSEP T-test of equal predictive ability of two models \citep[Section~2.3]{Rhoadesetal2011} is based on the information gain per earthquake,
\[
\mathrm{IG}_{(\hsp j,k)}^\mathrm{PE} = \frac{T}{N_T} \left( \bar\myS^{(\hsp j)} - \bar\myS^{(k)} \right) \! ,
\] 
of model $k$ over model $j$ as defined at \eqref{eq:IG}.  \citet{Rhoadesetal2011} posit that under the null hypothesis of equal predictive ability the statistic
\begin{equation}  \label{appeq:T}
\Theta_{(\hsp j,k)} = N_T^{1/2} \: \frac{\mathrm{IG}_{(\hsp j,k)}^\mathrm{PE}}{s_{(\hsp j,k)}}, 
\end{equation}
where 
\[
s_{(\hsp j,k)}^2 = 
\frac{1}{N_T - 1} \sum_{t = 1}^T \sum_{c = 1}^C y_{c,t} \! \left( \Delta_{c,t}^{(\hsp j,k)} \right)^2
- \frac{1}{N_T (N_T - 1)} \left( \sum_{t = 1}^T \sum_{c = 1}^C y_{c,t} \, \Delta_{c,t}^{(\hsp j,k)} \right)^{\! 2}
\]
and $\Delta_{c,t}^{(\hsp j,k)} = \log x_{c,t}^{(\hsp j)} - \log x_{c,t}^{(k)}$, has a Student's $t$-distribution with $N_T - 1$ degrees of freedom.  While $s_{(\hsp j,k)}$ aims to quantify the variability of the daily score differences, this line of reasoning suffers from the omission of terms associated to grid cells and time periods without target earthquakes, a neglect of the effects of spatio-temporal dependencies, and related issues in the estimation of sampling variability.\footnote{Specifically, the information gain per earthquake in Eq.~(17) of \citet{Rhoadesetal2011} is normalized by the observed count $N$ of target earthquakes.  Yet, it involves the quantities $\hat{N}_A$ and $\hat{N}_B$, i.e., data from all forecast bins regardless of occurred target earthquakes.  Hence, the quantity $N$ in the CSEP T-statistic in the line that follows Eq.~(18) of \citet{Rhoadesetal2011} ought to be replaced by the (typically much larger) number of all forecast bins, $n$, which is the proper sample size in this context (save for further adjustments in case of dependencies, e.g., from overlapping forecast windows). 

The sample variance in Eq.~(18) of \citet{Rhoadesetal2011} is only computed from $N$ bins (with strictly positive earthquake counts); instead, it ought to be computed from the Poisson score differences over all $n$ bins (regardless of occurred target earthquakes).  Furthermore, the sample variance involves the logarithmic terms in the Poisson score only, while ignoring the linear terms that sum to $\hat{N}_A$ and $\hat{N}_B$, respectively.  Additional detail is available from the authors upon request.

One might argue that the CSEP T-test attempts to estimate the variance of the information gain per earthquake (IGPE).  However, IGPE normalizes by the observed count $N$ of target earthquakes, which may be zero.  So, viewed as a random variable, IGPE may be undefined due to division by zero.  Fortunately, there is an easy way out of this conundrum, namely, to avoid division by the observed count and consider the information gain (IG) and score differences in lieu of IGPE, which in view of Eq.~\eqref{eq:IG} amounts to evolving the CSEP T-test into the Diebold--Mariano test.}  As a result, the test statistic \eqref{appeq:T} is not properly standardized, and in general the one-sided $p$ value generated by the T-test, namely, 
\begin{equation}  \label{appeq:p}
p = 1 - \Psi \! \left( \Theta_{(\hsp j,k)} \right) \! , 
\end{equation}
where $\Psi$ denotes the CDF of the Student's $t$-distribution with $N_T - 1$ degrees of freedom, fails to be uniform between 0 and 1 when the null hypothesis is true.

\begin{figure}[t]
\centering
\includegraphics[scale=1.0]{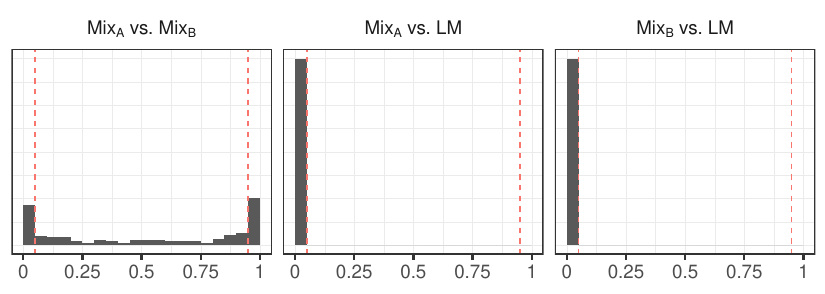}
\vspace{-2mm}
\caption{As Figure~\ref{fig:DM}, but for the CSEP T-test.  \label{fig:T}}
\end{figure}

\begin{table}[t]
\centering
\caption{As Table~\ref{tab:DM}, but for the CSEP T-test  of the null hypothesis of equal predictive ability.  We show the respective information gain per earthquake ($\mathrm{IG}_{(\hsp j,k)}^\mathrm{PE}$, diagonal) from Table~\ref{tab:IG}, the test statistic ($\Theta_{(\hsp j,k)}$, Eq.~\eqref{appeq:T}, above diagonal), and the one-sided $p$ value (Eq.~\eqref{appeq:p}, below diagonal).  Rows identify model $j$ and columns model $k$ in the test statistic at \eqref{appeq:T}.  \label{tab:T}}
\medskip
\begin{tabular}{lccccc}
\toprule
& \begin{turn}{90} LRWA \hsp \end{turn}
& \begin{turn}{90} LM \hsp \end{turn}
& \begin{turn}{90} SMA \hsp \end{turn}
& \begin{turn}{90} FCM \hsp \end{turn}
& \begin{turn}{90} LG \hsp \end{turn} \\
\midrule
LRWA &  \fbox{$-0.053$}  &   5.304   &   11.337   &   15.677   &   28.277  \\
LM   &   0.000   &  \fbox{ 0.000 }  &   4.380   &   11.382   &   23.965  \\
SMA  &   0.000   &   0.000   &  \fbox{ 0.061 }  &   10.641   &   30.186  \\
FCM  &   0.000   &   0.000   &   0.000   &  \fbox{ 0.253 }  &   18.942  \\
LG   &   0.000   &   0.000   &   0.000   &   0.000   &  \fbox{ 0.925 } \\
\bottomrule
\end{tabular}
\end{table}

We illustrate these issues in the simulation setting of Section~\ref{sec:DM}.  Figure~\ref{fig:T} is the same as Figure~\ref{fig:DM} in Section~\ref{sec:DM}, but now considering the CSEP T-test rather than the Diebold--Mariano test.  The left histogram in Figure~\ref{fig:T} shows the $p$ values that arise from 400 replicates of CSEP T-tests of Mix$_A$ (index $j$) versus Mix$_B$ (index $k$); we see that the histogram deviates considerably from the desired uniformity between 0 and 1.  Specifically, 85 of the 400 $p$ values are smaller than 0.05, and 101 of them are larger than 0.95, so that the two-sided CSEP T-test rejects at a ratio of $(85 + 101)/400 = 0.465$, grossly failing to attain the nominal level of 0.10 under the null hypothesis.  We conclude that the CSEP T-test rejects the null hypothesis of equal predictive ability more often than warranted.  This behavior is mirrored in Table~\ref{tab:T}, where the T-test rejects the null hypothesis of equal predictive ability for every pair of the five forecast models.

\begin{figure}[t]
\centering
\includegraphics[scale=1.0]{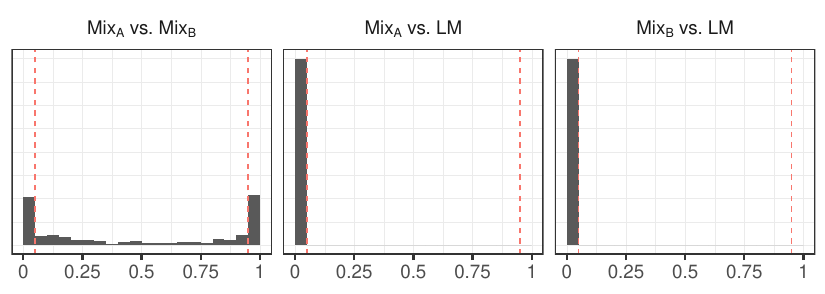}
\vspace{-2mm}
\caption{As Figure~\ref{fig:T}, but using forecasts on Mondays only.  \label{fig:T_Monday}}
\end{figure}

One might speculate that the undesirable behavior under the null hypothesis stems predominantly from the overlap of the seven-day forecast periods in OEF-Italy.  In this light, we repeat the analysis in Figure~\ref{fig:T_Monday} and Table~\ref{tab:T_Monday}, but now considering forecasts issued on Mondays only, to avoid overlap in the seven-day forecast periods.  Unfortunately, the aforementioned issues prevail.  In the 400 tests of Mix$_A$ versus Mix$_B$, 104 of the 400 $p$ values are smaller than 0.05, and 108 of them are larger than 0.95, rather than the nominal 20.

\begin{table}[t]
\centering
\caption{As Table~\ref{tab:T}, but using forecasts on Mondays only.  \label{tab:T_Monday}}
\medskip
\begin{tabular}{lccccc}
\toprule
& \begin{turn}{90} LRWA \hsp \end{turn}
& \begin{turn}{90} LM \hsp \end{turn}
& \begin{turn}{90} SMA \hsp \end{turn}
& \begin{turn}{90} FCM \hsp \end{turn}
& \begin{turn}{90} LG \hsp \end{turn} \\
\midrule
LRWA &  \fbox{$-0.068$} &   1.929   &   6.117   &   6.097   &   10.618  \\
LM   &   0.027   &  \fbox{ 0.000 }  &   1.777   &   4.650   &   8.549  \\
SMA  &   0.000   &   0.038   &  \fbox{ 0.070 }  &   3.964   &   10.664  \\
FCM  &   0.000   &   0.000   &   0.000   &  \fbox{ 0.266 }  &   6.731  \\
LG   &   0.000   &   0.000   &   0.000   &   0.000   &  \fbox{ 0.910 } \\
\bottomrule
\end{tabular}
\end{table}

\begin{figure}[ph]
\centering
\includegraphics[scale=0.975]{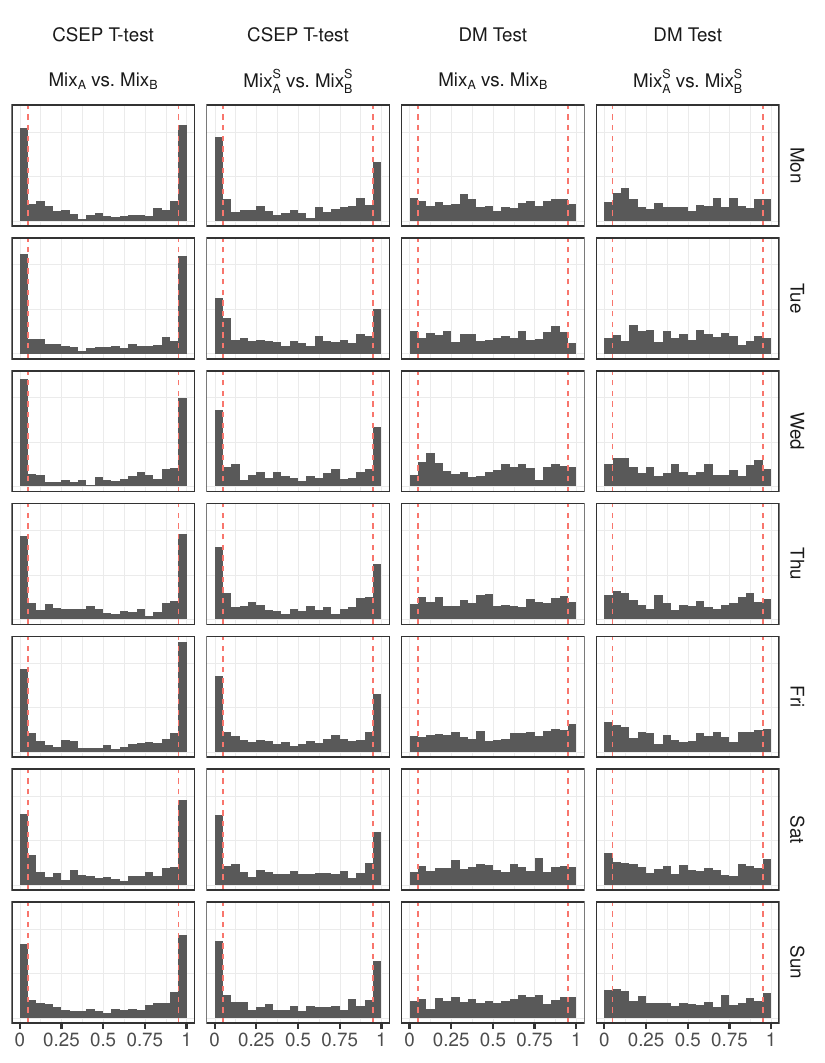}
\caption{Histograms of $p$ values for the CSEP T-test (Left) and the Diebold--Mariano (DM) test (Right) of equal predictive ability in terms of the Poisson scoring function for Mix$_A$ versus Mix$_B$, and Mix$_A^S$ versus Mix$_B^S$, respectively, based on 400 replicates.  The tests use forecasts for non-overlapping weekly periods issued Monday (Top) through Sunday (Bottom).  Compared to Figures~\ref{fig:DM}, \ref{fig:T}, and \ref{fig:T_Monday}, note the zoomed in vertical axis.  \label{fig:AB}}
\end{figure}

For a more comprehensive analysis see Figure~\ref{fig:AB}, where the first and third columns show histograms of $p$ values for the CSEP T-test and the Diebold--Mariano test of equal predictive ability for Mix$_A$ versus Mix$_B$ for non-overlapping weekly periods, issued Monday through Sunday, respectively.  While Mix$_A$ versus Mix$_B$ randomize between the FCM model and the LG model at any given time $t$, another option is to randomize between the two models at any grid cell $c$, with the assignment kept fixed over time.  We refer to the spatially randomized forecasts as Mix$_A^S$ and Mix$_B^S$, respectively, and show histograms of $p$ values for the CSEP T-test and Diebold--Mariano test of equal predictive ability for Mix$_A^S$ versus Mix$_B^S$ in the second and fourth column of Figure~\ref{fig:AB}.  The Diebold--Mariano test always yields uniform histograms for the $p$ values, as desired.  The CSEP T-test instead yields U-shaped histograms, resulting from an underestimated variance of score differences, which entails test statistics that are larger (in absolute value) than warranted and, consequently, rejections of the null hypothesis well beyond the nominal level.

\section{Forecast performance at the grid cell level}  \label{app:individual}

\begin{figure}[p]
\centering
\includegraphics[scale=0.99]{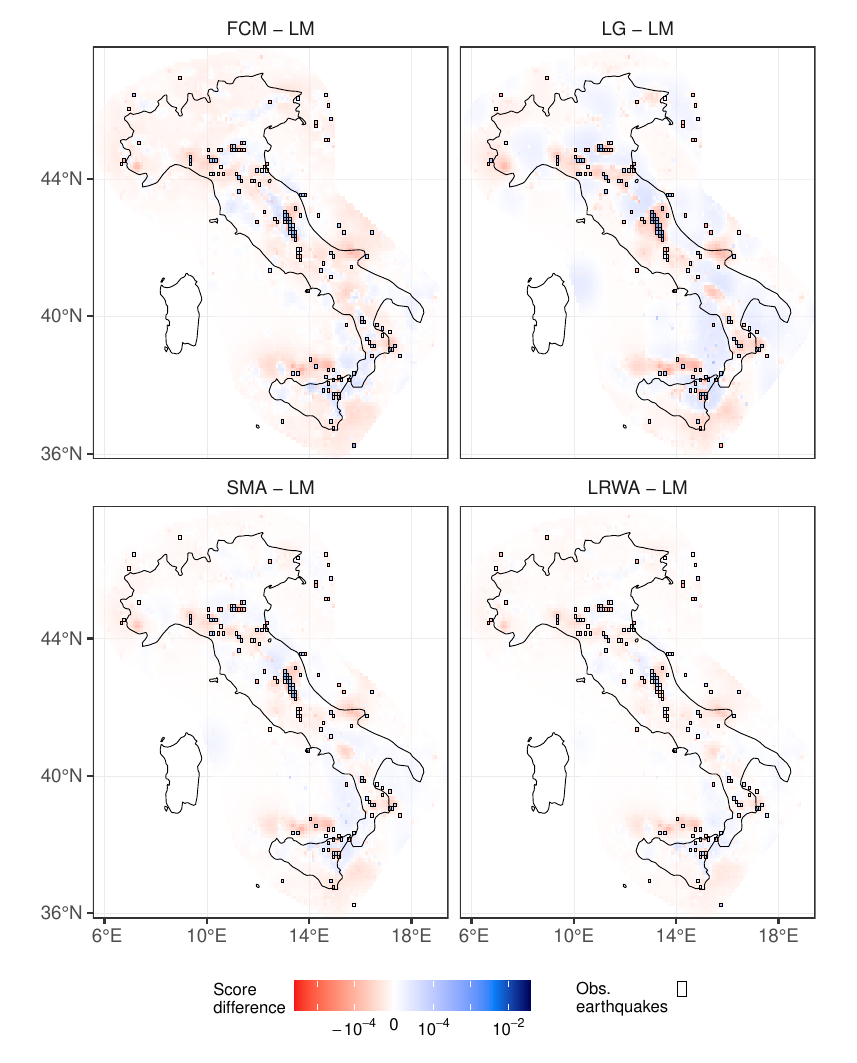}
\caption{Average score difference \eqref{eq:score_diff} under the Poisson scoring function between the LM (index $k$) and the other four models (index $j$, shown panels).  Framed cells represent bins with at least one target earthquake.  Positive differences (blue color) indicate that the LM model has superior forecast performance; negative differences (red color) indicate inferior performance.  \label{fig:spatial_comp}}
\end{figure}

While Sections \ref{sec:consistent} and \ref{sec:DM} considered overall and temporal assessments of a model's predictive performance, respectively, we might also ask in which areas of the testing region one can find remarkable differences in forecast performance.  To do so, we can study the performance via temporally aggregated scores for each of the 8993 grid cells in OEF-Italy.  Specifically, we consider the average score difference between model $j$ and model $k$ in grid cell $c$, given by
\begin{equation}  \label{eq:score_diff}
\Delta_c^{(\hsp j,k)} = \frac{1}{T} \sum_{t = 1}^T \left( \myS(x_{c,t}^{(\hsp j)}, y_{c,t}) - \myS(x_{c,t}^{(k)}, y_{c,t}) \right) \! ,
\end{equation}
where $\myS$ is the Poisson scoring function.  We here use the average instead of the total score to normalize to daily performance.  We again let $k$ stand for the LM model.  Positive values of $\Delta_c^{(\hsp j,k)}$ indicate that the LM model produces superior forecasts in grid cell $c$.

Figure~\ref{fig:spatial_comp} plots $\Delta_c^{(\hsp j,k)}$, i.e., the average score difference between the LM model and model $j$ in grid cell $c$, for the four competitor models.  In line with our previous analysis, the LM model exhibits superior forecast performance compared to FCM, SMA, and LM in most grid cells with observed M4+ target earthquakes, especially in central Italy.  While the FCM model receives lower scores than the LM model in regions with less seismicity, the LG model performs worse than the LM model in those.  Compared to LRWA, the sign of the score differences is more mixed.  We emphasize that these results are to be interpreted diagnostically, with formal statistical inference being challenging in spatial settings due to massive dependencies and issues of multiple testing \citep{Zhang2015, Wilks2016}.

\section{Generation of consistency bands for mean-reliability diagrams}  \label{app:consistency_bands}

To accommodate for the non-negative integer character of earthquake counts we use a resampling procedure based on Algorithm 3 in \citet{GneitResin2022} for the generation of the consistency bands in Figure~\ref{fig:reliability_CSEP}.  The algorithm requires fully specified predictive distributions from which it then samples.  In the setting at hand we have mean-forecasts (more specifically, expected counts) at our disposal only.  To derive a fully specified predictive distribution from the mean-forecast at hand, we adapt the unconditional distribution of the number of observed target earthquakes.  Specifically, if the vector $(p_0, p_1, \ldots, p_m) \in [0,1]^{m+1}$ comprises the empirical frequencies of observing $0, 1, \ldots, m$ earthquakes in the record at hand, we derive the predictive distribution $F_x$ associated with the mean-forecast $x$ as the discrete probability measure on $0, 1, \ldots, m$ with masses $(p_0 + \varepsilon_x, p_1, \ldots, p_m) / (1 + \varepsilon_x)$, where $\varepsilon_x$ is such that 
\begin{equation}  \label{appeq:epsilon}
x = \frac{1}{1 + \varepsilon_x} \sum_{j = 1}^m j \, p_j,
\end{equation}
which ensures that the distribution $F_x$ has mean $x$.  Then we apply Algorithm 3 in \citet{GneitResin2022} to the respective collection of pairs $(F_x, x)$. 

\end{document}